\documentclass[a4paper,11pt]{article}\usepackage[]{graphicx}\usepackage[]{color}
\makeatletter
\def\maxwidth{ %
  \ifdim\Gin@nat@width>\linewidth
    \linewidth
  \else
    \Gin@nat@width
  \fi
}
\makeatother

\definecolor{fgcolor}{rgb}{0.345, 0.345, 0.345}
\newcommand{\hlnum}[1]{\textcolor[rgb]{0.686,0.059,0.569}{#1}}%
\newcommand{\hlstr}[1]{\textcolor[rgb]{0.192,0.494,0.8}{#1}}%
\newcommand{\hlcom}[1]{\textcolor[rgb]{0.678,0.584,0.686}{\textit{#1}}}%
\newcommand{\hlopt}[1]{\textcolor[rgb]{0,0,0}{#1}}%
\newcommand{\hlstd}[1]{\textcolor[rgb]{0.345,0.345,0.345}{#1}}%
\newcommand{\hlkwa}[1]{\textcolor[rgb]{0.161,0.373,0.58}{\textbf{#1}}}%
\newcommand{\hlkwb}[1]{\textcolor[rgb]{0.69,0.353,0.396}{#1}}%
\newcommand{\hlkwc}[1]{\textcolor[rgb]{0.333,0.667,0.333}{#1}}%
\newcommand{\hlkwd}[1]{\textcolor[rgb]{0.737,0.353,0.396}{\textbf{#1}}}%

\usepackage{framed}
\makeatletter
\newenvironment{kframe}{%
 \def\at@end@of@kframe{}%
 \ifinner\ifhmode%
  \def\at@end@of@kframe{\end{minipage}}%
  \begin{minipage}{\columnwidth}%
 \fi\fi%
 \def\FrameCommand##1{\hskip\@totalleftmargin \hskip-\fboxsep
 \colorbox{shadecolor}{##1}\hskip-\fboxsep
     \hskip-\linewidth \hskip-\@totalleftmargin \hskip\columnwidth}%
 \MakeFramed {\advance\hsize-\width
   \@totalleftmargin\z@ \linewidth\hsize
   \@setminipage}}%
 {\par\unskip\endMakeFramed%
 \at@end@of@kframe}
\makeatother

\definecolor{shadecolor}{rgb}{.97, .97, .97}
\definecolor{messagecolor}{rgb}{0, 0, 0}
\definecolor{warningcolor}{rgb}{1, 0, 1}
\definecolor{errorcolor}{rgb}{1, 0, 0}
\newenvironment{knitrout}{}{} 

\usepackage{alltt}
\usepackage[scale={0.8,0.9},centering,includeheadfoot]{geometry}
\usepackage{authblk}
\usepackage[normalem]{ulem}

\title{Integrated Nested Laplace Approximations (INLA)}

\author{Sara Martino}
\author{Andrea Riebler}

\affil{Department of Mathematical Sciences,
    Norwegian University of Science and Technology,
    Norway}

\usepackage[numbers, super]{natbib}
\usepackage{amsthm} 
\usepackage{amsfonts} 
\usepackage{amscd} 
\usepackage{amsmath} 
\usepackage{graphicx}
\usepackage{epstopdf}
\usepackage{pgf}
\usepackage{array, multirow}
\usepackage{caption}
\usepackage{subcaption}
\usepackage{bm}
\usepackage{ltablex}
\usepackage{tabularx}

\usepackage{booktabs}

\usepackage{hyperref}

\usepackage[utf8]{inputenc}

\IfFileExists{upquote.sty}{\usepackage{upquote}}{}
\begin{document}

\maketitle

\begin{abstract}
This is a short description and basic introduction to the Integrated nested
Laplace approximations (INLA) approach. INLA is a deterministic
paradigm for Bayesian inference in latent Gaussian models (LGMs) introduced in \citet{rue2009}. 
INLA relies on  a combination of analytical approximations and
efficient numerical integration schemes to achieve highly accurate
deterministic approximations to posterior quantities of interest. The
main benefit of using INLA instead of Markov chain Monte Carlo (MCMC)
techniques for LGMs is computational; INLA is fast even for large,
complex models. Moreover, being a deterministic algorithm, INLA does
not suffer from slow convergence and poor mixing. INLA is implemented
in the R package {\tt R-INLA}, which represents a user-friendly and
versatile tool for doing Bayesian inference. {\tt R-INLA} returns posterior
marginals for all model parameters and the corresponding posterior summary
information. Model choice criteria as well as predictive diagnostics
are directly available. Here, we outline the theory behind INLA,
present the {\tt R-INLA} package and describe new developments of
combining INLA with MCMC for models that are not possible to fit with {\tt
  R-INLA}.
\end{abstract}


\section{Where can INLA be applied}\label{Section:LGM}

Latent Gaussian Models (LGM) are the class of Bayesian models amenable
to INLA-based inference. An LGM consists of three elements: a likelihood
model, a latent Gaussian field and a vector of hyperparameters. The
data $\bm{y}$ are assumed to be conditionally independent given the
latent Gaussian field $\bm{x}$ so that the univariate likelihood model
describes the marginal distribution of the observation. As in the
generalized linear model framework, the mean, or another measure of
central tendency, of the observation $y_i$ is linked to a Gaussian linear predictor $\eta_i$ through a known link function. The linear predictor is then additive with respect to other effects:
\begin{equation}\label{eq:lp}
  \eta_i = \mu + \sum_{j}\beta_jz_{ij} +\sum_{k}w_k\ f^{k}(u_{ik})
\end{equation}
Here $\mu$ is an overall intercept, $\bm{z}$ are known covariates
with linear effect $\bm{\beta}$ and $\bm{w}$ a vector of  known weights. The terms $\bm{f}^k$ are used to
model random effects of the covariate $\bm{u}$.
In the INLA framework we assign
 $\mu$,  $\bm{\beta}$  and $\bm{f}^k$, $k = 1,\dots,K$ a Gaussian prior. The latent Gaussian field is then
$\bm{x} = \{\bm{\eta}, \mu, \bm{\beta}, \bm{f}^1,\bm{f}^2,\dots\}$.
 Note that we include the linear predictor in the latent field. 
This is mainly
due to the fact having each data point $y_i$ dependent on the latent
Gaussian field only through one single element of $\bm{x}$, namely
$\eta_i$ greatly simplifies the computations needed in the INLA
algorithm, see for example \citet{rue2009} and \citet{inla_review1}
for details. For this reason, a small random noise term
where the precision parameter is fixed to a high value 
is always automatically added to the model.

The hyperparameters  $\bm{\theta}$ can appear in the likelihood model (for example the variance in the Gaussian likelihood or the shape parameter in the Gamma one) or/and in the latent field, typically as dispersion parameters, spatial correlation parameters or autoregression coefficients in the  $\bm{f}^k$ terms.  Formally the model can be written as:
\begin{align*}
\bm{y}|\bm{x},\bm{\theta}&\sim\prod \pi(y_i|\eta_i,\bm{\theta})\\
\bm{x}|\bm{\theta}&\sim \mathcal{N}(\bm{0},\bm{Q}^{-1}(\bm{\theta}))\\
\bm{\theta}&\sim\pi(\bm{\theta})
\end{align*}
where $\bm{Q}(\bm{\theta})$ is the precision (inverse of the
covariance) matrix of the latent Gaussian field.
A limitation of the INLA approach is the size of the hyperparameter vector $\bm{\theta}$. While $\bm{y}$ and $\bm{x}$ can be large, $\bm{\theta}$ should be small, say $<15$. This is due to the numerical integration that has to be carried over the $\bm{\theta}$ space. The dependence structure of the data is mainly captured by the precision matrix $\bm{Q}(\bm{\theta})$ through a clever choice of the terms $\bm{f}^k$ in \eqref{eq:lp}.
In order for INLA to perform efficiently, we require the precision matrix $\bm{Q}(\bm{\theta})$ to be sparse.

Many of the models that are commonly used as prior for the $\bm{f}^k$ terms belong to the class of
so called Gaussian Markov Random field (GMRF). GMRFs can be used to  model
smooth effects of a covariate, random effects, measurement errors,
temporal dependencies, and so on (see \citet{Rue_Held_book2005}).
When it comes to spatial dependence, there exist GMRF models for areal
data, such as the CAR or BYM model proposed by \citet{bym}. Continuous
spatial dependence can be specified using the so-called SPDE approach
\citep{SPDE-paper, bakka-etal-2018} which creates an approximated  GMRF representation
of the Mat{\'e}rn covariance field based on stochastic partial
differentiation equations.
GMRFs are Gaussian models endowed with Markov properties. These, in turn, are
linked to the non-zero structure of the precision matrix in the sense
that, if two elements of the field are conditionally independent given
all the others, then the corresponding entry of the precision matrix
is equal to zero, see \citet{Rue_Held_book2005}. In practice, choosing GMRF priors for $\bm{f}^k$, induces sparsity in the  precision matrix $\bm{Q}(\bm{\theta})$.

The resulting posterior density of $\bm{x}$ and $\bm{\theta}$ given $\bm{y}$ is:
\begin{equation}
\pi(\bm{x}, \bm{\theta}|\bm{y})\propto\exp\left(-\frac{1}{2}\bm{x}^T \bm{Q}(\bm{\theta})\bm{x} + \sum_i\log(\pi(y_i|\eta_i,\bm{\theta}))+\log\pi(\bm{\theta})
\right)
\end{equation}
This is a high dimensional density that is difficult to
interpret. Often the main interest lies in the marginal posterior of
the latent field $\pi(x_i|\bm{y})$ or of the hyperparameters
$\pi(\theta_j|\bm{y})$.
INLA provides an approximation to such marginal posterior densities,
which can then be used to compute approximated summary statistics of interest such as posterior means, variances or quantiles.

To sum up, INLA can be applied to LGMs which fulfill the following assumptions:
\begin{enumerate}

  \item Each data point depends on only one of the elements in the latent
    Gaussian field $\bm{x}$, the linear predictor,  so that the likelihood
    can be written as:
    \[
       \bm{y}|\bm{x},\bm{\theta}\sim\prod_{i}\pi(y_i|\eta_i,\bm{\theta}).
     \]
  \item The size of the hyperparameter vector $\bm{\theta}$ is small
    (say $<15$)
  \item The latent field $\bm{x}$, can be large but it  is endowed with some conditional
    independence (Markov) properties so that the precision matrix
    $Q(\bm{\theta})$ is sparse.
    \item The linear predictor depends linearly on the unknown smooth function of covariates.
    \item The inferential interest lies in the univariate posterior marginals
      $\pi(x_i|\bm{y})$ and $\pi(\theta_j|\bm{y})$ rather than in the
      joint posterior $\pi(\bm{x},\bm{\theta}|\bm{y})$.
\end{enumerate}

There are some, rather extreme, cases of LGM where the use of INLA has
been seen as problematic \citep{ferkingstad2015, SauterHeld2016}. If the likelihood is binomial or Poisson, inaccuracies can occur with low counts and low degree of smoothing. Variances of both random and fixed effects tend to be underestimated by INLA while means are usually well estimated. \citet{ferkingstad2015} introduce a correction, implemented in the {\tt R-INLA} package, that, at a negligible computational cost, alleviates the problem.

\section{The INLA computing scheme}\label{Sec:comp_scheme}

The INLA framework provides deterministic approximations  to the
univariate posterior marginals for the hyperparameters
$\pi(\theta_j|\bm{y})$ and the latent field $\pi(x_i|\bm{y})$. Thus, interest lies in:
\begin{align}
        \label{eq.integrals1}
        \pi(\theta_j|\bm{y}) & =  \int \int
                               \pi(\bm{x},\bm{\theta}|\bm{x}) d\bm{x}
                               \ d\bm{\theta}_{-j}
                              = \int \pi(\bm{\theta}|\bm{y})\ d
                               \bm{\theta}_j
  \\
           \label{eq.integrals2}
        \pi(x_i|\bm{y}) & = \int \int
                               \pi(\bm{x},\bm{\theta}|\bm{x}) d\bm{x}_{-i}
                               \ d\bm{\theta}  = \int \pi(x_i|\bm{\theta},\bm{y}) \pi(\bm{\theta}|\bm{y}) \ d \bm{\theta}
\end{align}
Notice that, because of the model requirements explained in Section
\ref{Section:LGM} the integral with respect to $\bm{x}$ in \eqref{eq.integrals1} and
\eqref{eq.integrals2}  can be (and usually is) highly multidimensional,
while the  integral with respect to $\bm{\theta}$ is only moderate in
size and can be solved via some numerical integration scheme. The core
of the INLA methodology lies therefore in building clever
approximations to the posterior for the hyperparameters
$\pi(\bm{\theta}|\bm{y}) $ and the full-conditional density
$\pi(x_i|\bm{\theta},\bm{y})$ that allow to avoid the cumbersome integration with

An approximation to $\pi(\bm{\theta}|\bm{y}) $ is built starting from
the identity:
\begin{equation}\label{eq:theta}
          \pi(\bm{\theta}|\bm{y}) =
   \frac{\pi(\bm{x},\bm{\theta}|\bm{y})} {\pi(\bm{x}|\bm{\theta},\bm{y})}\propto
   \frac{\pi(\bm{y}| \bm{x}, \bm{\theta}) \pi(\bm{x}|\bm{\theta})\pi(\bm{\theta})}{\pi(\bm{x}|\bm{\theta},\bm{y})}
\end{equation}
Notice that, while the numerator  in \eqref{eq:theta} is easy to
compute, the denominator is, in general, not available in closed form and hard
to compute. INLA approximates \eqref{eq:theta} at a specific value
$\bm{\theta}^k$ of the hyperparameters vector as:
\begin{equation}\label{eq:theta1}
          \widetilde{\pi}(\bm{\theta}^k|\bm{y})\propto
   \frac{\pi(\bm{y}| \bm{x}, \bm{\theta}^k) \pi(\bm{x}|\bm{\theta}^k)\pi(\bm{\theta}^k)}{\widetilde{\pi}_G(\bm{x}|\bm{\theta}^k,\bm{y})}
\end{equation}
where $\widetilde{\pi}_G(\bm{x}|\bm{\theta}^k,\bm{y})$ is a Gaussian approximation to the full conditional
$\bm{x}|\bm{\theta}^k,\bm{y}$ build by matching the mode and the curvature at the mode. This expression is equivalent to \citet{Tierney_Kadane}'s Laplace approximation of a marginal posterior distribution.
The computationally expensive part of evaluating
\eqref{eq:theta1} is the Cholesky decomposition of the
$Q(\bm{\theta}^k)$ matrix necessary to evaluate the denominator and that
needs to be performed for each value $\bm{\theta}^k$. Here, the
sparseness of the precision matrix is essential, see \citet{rue2009} for more details.

Next, we need to find an approximation
$\widetilde{\pi}(x_i|\bm{\theta}^k,\bm{y})$  of
$x_i|\bm{\theta}^k,\bm{y}$. This step is more involved as it has to be
repeated for each element of the, virtually very large dimensional,
vector $\bm{x}$. 
One could use the marginal from the Gaussian approximation $\widetilde{\pi}_G(\bm{x}|\bm{\theta}^k,\bm{y})$ from \ref{eq:theta1}. While this is very fast, it is usually not very accurate.
As an alternative, we can start by writing $\pi(x_i|\bm{\theta}^k,\bm{y})$ as
\begin{equation}\label{eq:x}
\pi(x_i|\bm{\theta}^k,\bm{y})=
  \frac{\pi(\bm{x}|\bm{\theta},\bm{y})} {\pi(\bm{x}_{-i}|x_i,\bm{\theta},\bm{y})}\propto
   \frac{\pi(\bm{y}| \bm{x}, \bm{\theta}) \pi(\bm{x}|\bm{\theta})\pi(\bm{\theta})}{\pi(\bm{x}_{-i}|x_i,\bm{\theta},\bm{y})}
\end{equation}
where $\bm{x}_{-i}$ indicates the vector $\bm{x}$ without the $i^{\text{th}}$ element. This expression is similar to \ref{eq:theta}.  
An approximation to $\pi(x_i|\bm{\theta}^k,\bm{y})$
can then be constructed by approximating the denominator in \eqref{eq:x} by matching the mode and the curvature at the mode. This is again equivalent to \citet{Tierney_Kadane}'s Laplace approximation. The problem with \eqref{eq:theta} is that it is very computationally demanding as it requires factorizing many times a large precision matrix.
\citet{rue2009} propose therefore a third approximation, denoted the Simplified Laplace approximation, which corrects the Gaussian approximation for location and skewness by a Taylor's series expansion around the mode of the Laplace approximation. All three approximations are available in the {\tt R-INLA} package. The Simplified Laplace is the default choice.

The last step is the numerical integration scheme to solve the
integral with respect to $\bm{\theta}$ in \eqref{eq.integrals1} and
\ref{eq.integrals2}. The  {\tt R-INLA} package offers three possible
alternatives. The first is to build a grid on the $\bm{\theta}$ space
around the mode of $\widetilde{\pi}(\bm{\theta}|\bm{y})$ ({\tt int.strategy='grid'}). This
strategy gives the most accurate approximations but the number of
points in the grid grows exponentially with the size of the
$\bm{\theta}$ vector. It is the default choice in the  {\tt
  R-INLA} package if the dimension of $\bm{\theta}$ is one or two.
The second strategy is to use a so called
central composite design to cleverly locate fewer points around the
mode of $\widetilde{\pi}(\bm{\theta}|\bm{y})$ ({\tt
  int.strategy='ccd'}). This is the default strategy for dimensions of $\bm{\theta}$ larger than two. Finally one can ignore the variability of the
hyperparameter and just use the mode of
$\widetilde{\pi}(\bm{\theta}|\bm{y})$ ({\tt
  int.strategy='eb'}).

Putting all together, the INLA computing scheme is as follows:

\begin{enumerate}
   \item Explore the $\bm{\theta}$ space through the approximation
   $\widetilde{\pi}(\bm{\theta}|\bm{y})$.  Find the mode of
     $\widetilde{\pi}(\bm{\theta}|\bm{y})$ and locate a series of points
       $\{\bm{\theta}^1,\dots, \bm{\theta}^K\}$ in the area of high
       density of  $\widetilde{\pi}(\bm{\theta}|\bm{y})$.
 \item For the $K$ selected support points compute
 $\widetilde{\pi}(\bm{\theta}^1|\bm{y}), \ldots,\widetilde{\pi}(\bm{\theta}^K|\bm{y}) $ using \eqref{eq:theta1}.
\item  For each selected $\bm{\theta}^k$ point, approximate the
  density of $x_i|\bm{\theta},\bm{y}$ as
  $\widetilde{\pi}(x_i|\bm{\theta}^k,\bm{y})$ for $k = 1,\dots,K$
  using one of the three possible approximations: Laplace,
  Simplified Laplace or Gaussian.
\item Solve \eqref{eq.integrals2} via numerical integration as:
\begin{equation}\label{eq:INLA_int}
        \widetilde{\pi}(x_i|\bm{y}) = \sum_{k = 1}^K \widetilde{\pi}(x_i|\bm{\theta}^k,\bm{y}) \widetilde{\pi}(\bm{\theta}^k|\bm{y})\Delta_k
\end{equation}
Where $\Delta_k$ are appropriate weights, which would be equal to 1
for example if all support points would be equi-distantly chosen. The integral \eqref{eq.integrals1} can be solved similarly.
\end{enumerate}
The scheme above sheds light on the name INLA: the nested Laplace approximations are those performed in steps 2 and 3 while the integrated bit comes from the numerical integration in step 4.

Note that the error committed in \eqref{eq:INLA_int} comes from two different sources: one
is the approximation error due to approximating
$\pi(x_i|\bm{\theta}^k,\bm{y})$ with
$\widetilde{\pi}(x_i|\bm{\theta}^k,\bm{y})$,
the other is due to the numerical integration scheme and the choice of
the support points $\bm{\theta}^k$. Using a Gaussian likelihood, the full
conditional $\pi(\bm{x}|\bm{\theta},\bm{y})$ and (of course) its
marginals  $\pi(x_i|\bm{\theta},\bm{y})$ are also Gaussian. This
implies that, for each value of $\bm{\theta}$,  \eqref{eq:theta} can be
computed exactly. The only source of error in \eqref{eq:INLA_int} is
then the numerical integration scheme. We will look in details at this
special case in the Section \ref{Section1}.

An interesting feature of INLA is that it can approximate, as a bi-product
of the main computations, leave-one-out cross-validatory model checks
without re-running the model with individually removed
observations. These include the conditional predictive ordinate (CPO)
and probability integral transform (PIT) values that can be used to
asses the quality of the model. See
\citet{crossval2010} for details about how these measures are computed
and a comparison with MCMC results. The {\tt R-INLA} package returns
an additional flag vector indicating when an observation-specific CPO
values is not accurately approximated, and offers the user the helper
function {\tt inla.cpo} to replace
this value with the correct value obtained by removing the corresponding observation from the data frame and
refitting the model. INLA can also provide estimates for deviance information criterion (DIC) \citep{DIC2002}, Watanabe-Akaike information criterion (WAIC) \citep{Watanabe:2010} and marginal likelihood. The marginal likelihood is a well established model selection criterion in Bayesian statistics and can be used for Bayesian model averaging. Recently \citet{Hubin2016} have studied the accuracy of the marginal likelihood estimate provided by INLA finding it very accurate.

\section{The Gaussian Likelihood case}\label{Section1}

\begin{figure}[h]
  \begin{center}
    \begin{subfigure}[b]{0.48\textwidth}
      \includegraphics[width=\textwidth]{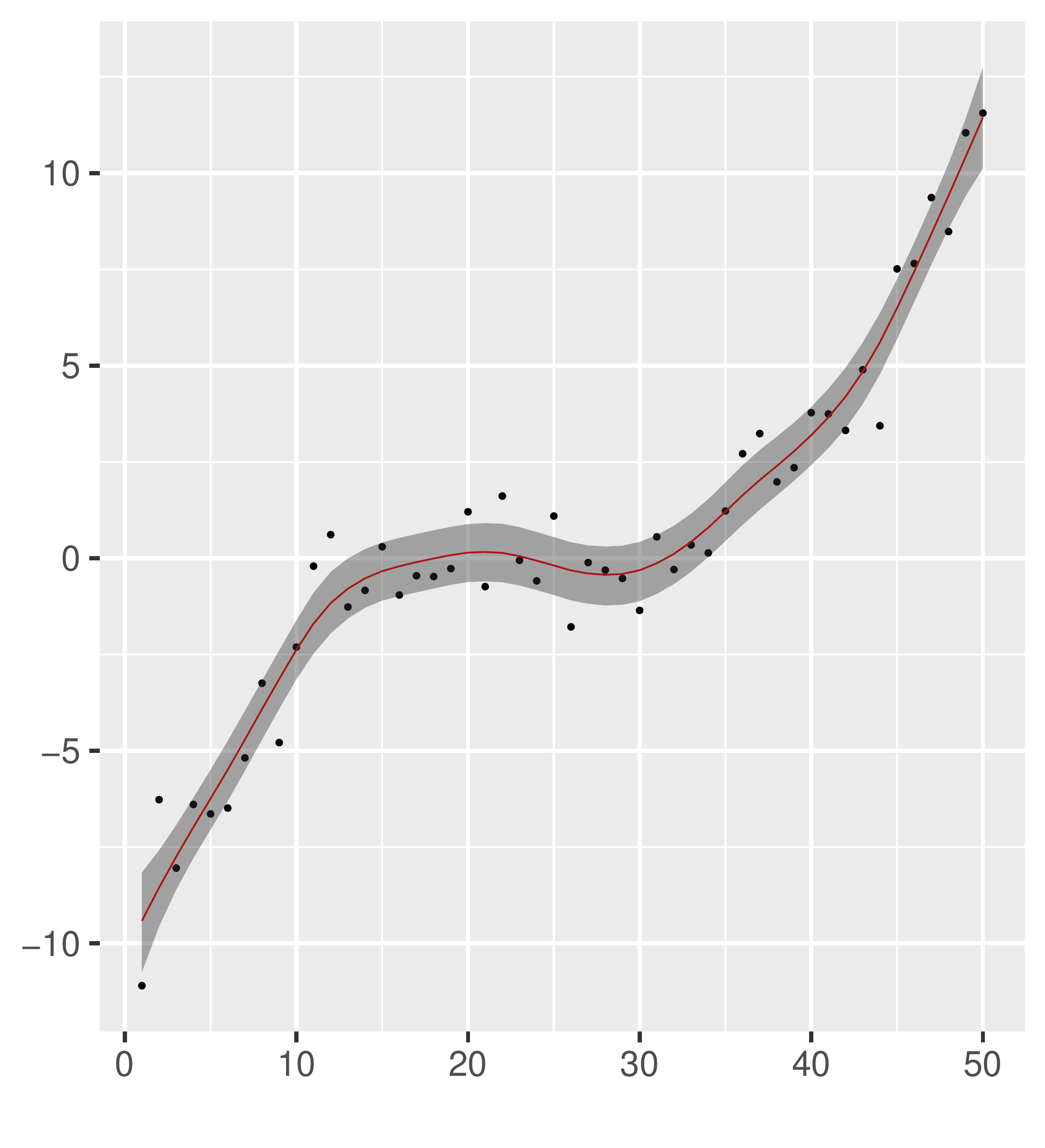}
      \caption{}\label{Fig:ex1_1}
    \end{subfigure}
    \begin{subfigure}[b]{0.48\textwidth}
      \includegraphics[width=\textwidth]{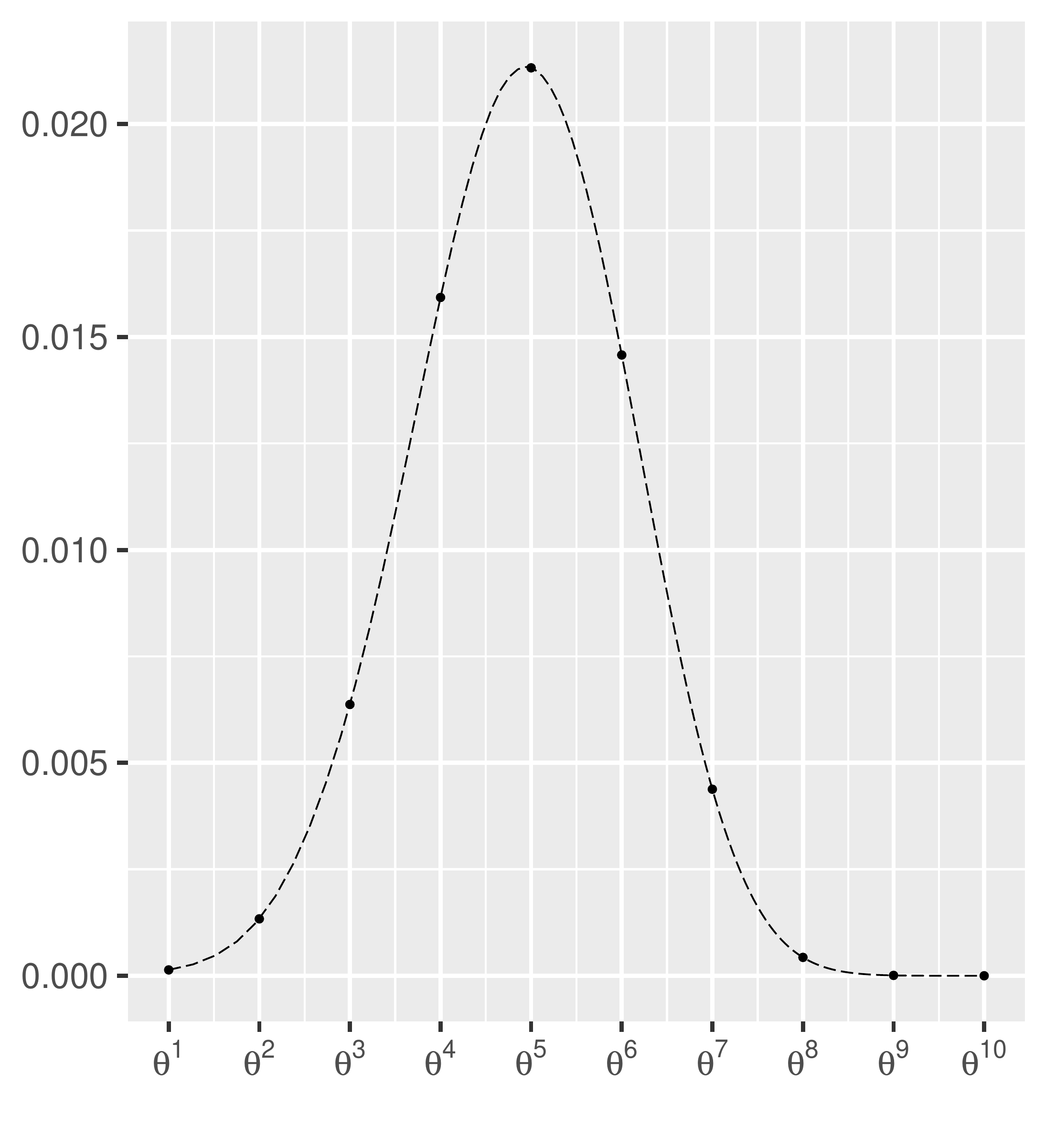}
      \caption{}\label{Fig:ex1_2}
    \end{subfigure}
            \vskip\baselineskip

     \begin{subfigure}[b]{0.48\textwidth}
      \includegraphics[width=\textwidth]{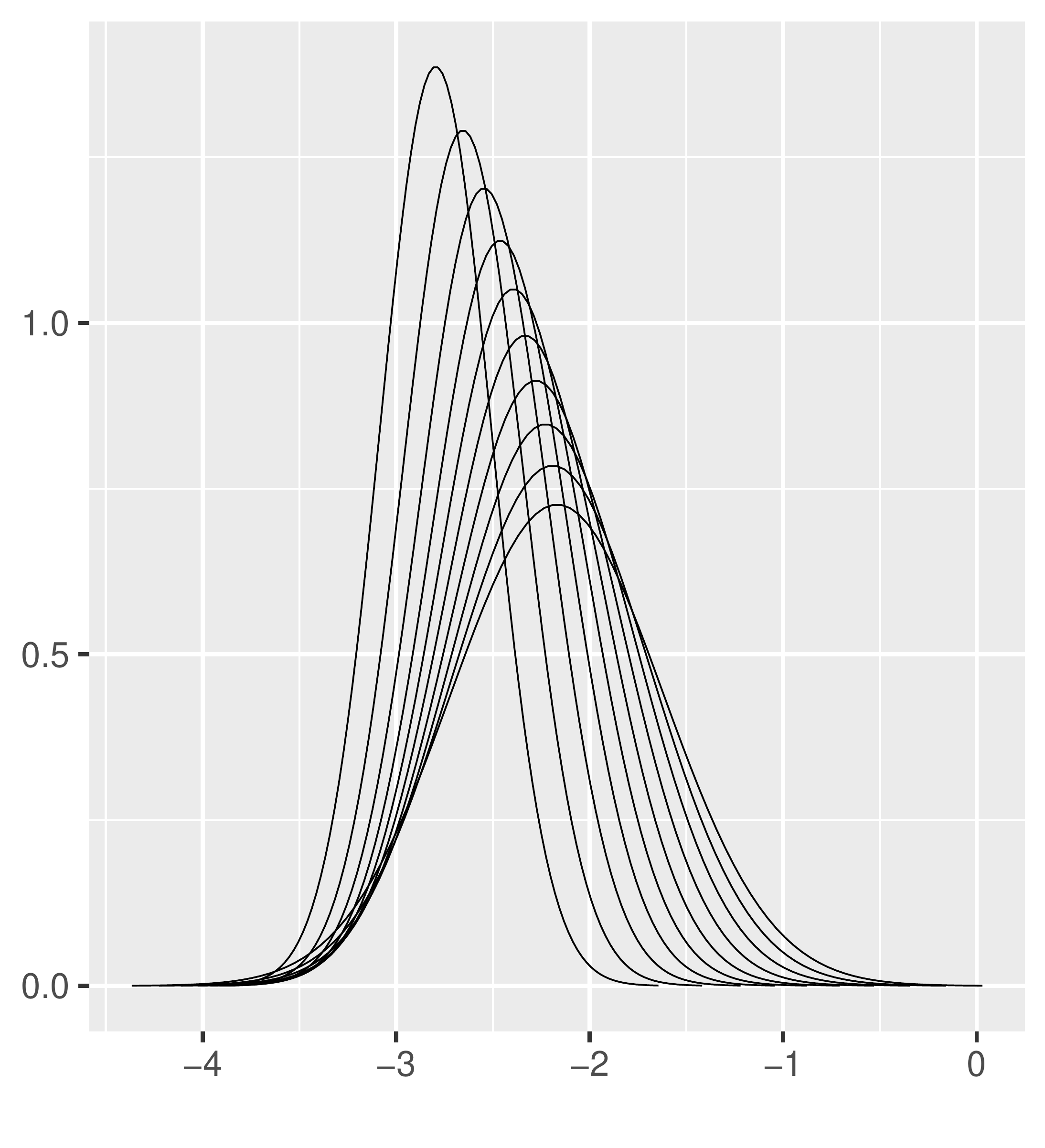}
      \caption{}\label{Fig:ex1_3}
    \end{subfigure}
    \begin{subfigure}[b]{0.48\textwidth}
      \includegraphics[width=\textwidth]{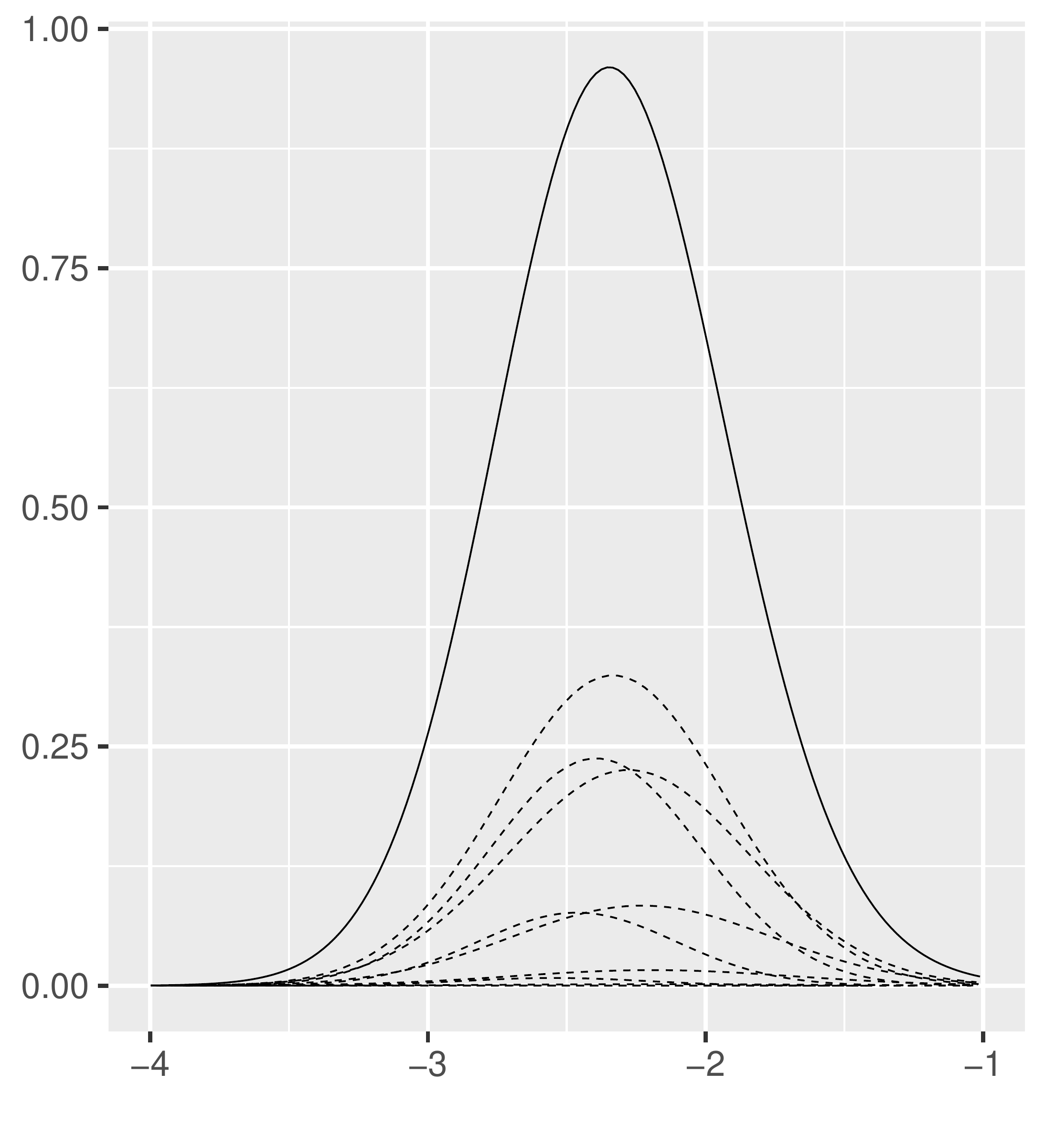}
      \caption{}\label{Fig:ex1_5}
    \end{subfigure}
  \end{center}
    \caption{ a) Observed
      time series (dots) together with the posterior estimated
      mean (black line). The grey band indicates a 95\%
      pointwise credible interval around the estimated smooth curve.  b) Posterior distribution for the hyperparameter
      $\pi(\theta|\bm{y})$. The black
        dotes indicate the density at the chosen points
      $\theta^1,\dots,\theta^K$. c) unweighted full conditional densities
      $\pi(x_{10}|\bm{y},\theta^k)$, $k = 1,\dots,K$. d) Broken lines: full conditional densities
      $\pi(x_{10}|\bm{y},\theta^k)$, weighed by
      $\pi(\theta^k|\bm{y})\Delta_k$,  $k =
      1,\dots,K$. The solid line indicates the approximation $\tilde{\pi}(x_{10}|\bm{y})$ obtained by summing the weighted full conditional densities. }
  \end{figure}

Assume we observe the time series shown as dots in Figure~\ref{Fig:ex1_1}
and our goal is to recover the underlying smooth trend. Assume
that, given the vector $\bm{\eta} = (\eta_1,\dots,\eta_T)$, the
observations $y_t$ are independent and Gaussian distributed with mean
$\eta_t$ and known unit variance:
\[
  y_t|\eta_t = \mathcal{N}(\eta_t,1);\ t = 1,\dots,T
\]
The linear predictor $\eta_t$ is linked to a smooth effect of time $t$
as:
\[
  \eta_t = f(t)
\]

Random walk models are a popular choice for modeling smooth effects of
covariates or, as in this case, temporal effects (see for example
chapter 3 in \citet{Rue_Held_book2005}). Here, we choose a second order
random walk model as prior distribution for the vector ${\bm f} =
(f(1), \dots,f(T))$, so that:
\[
\pi({\bm f}|\theta)\propto \theta^{(T-2)/2}\exp\left\{
                    -\frac{\theta}{2}\sum_{t=3}^T[f(t) - 2f(t-1) + f(t-2)]^2
                    \right\} = \mathcal{N}(\bm{0},\mathbf{Q}({\theta})^{-1}).
\]
Thus, ${\bm f}|\theta$ is Gaussian distributed with mean $\bm{0}$ and
precision (inverse covariance) matrix $\mathbf{Q}({\theta})$. The
precision parameter
$\theta$ controls the smoothness of the vector $\bm{f}$. Note that the
precision matrix $\mathbf{Q}(\theta)$ is a band matrix with bandwidth $2$ and
therefore it is sparse.
We complete the model by assigning $\theta$ a prior
distribution. Here, we choose the newly proposed penalised complexity
(PC) prior \citep{simpson2017} with parameters $u = 1$ and $\alpha =
0.01$. This is equivalent to using an exponential prior with mean
$-u/\log(\alpha)$ for the standard deviation parameter $1/\sqrt{\theta}$.
The resulting model is an LGM that fulfills the requirements
listed earlier, namely: each data point depends on only one
element of the latent field,  the precision matrix of the latent
Gaussian field is sparse and we have only one hyperparameter.
Our main inferential interest lies in the posterior marginal for the
smooth effect $\pi(f(t)|\bm{y})$,  $t = 1,\dots,T$.
We follow the scheme outlined in Section \ref{Sec:comp_scheme}:
After finding the mode of $\theta|\bm{y}$ via an optimization
algorithm, we select  support points $\{\theta^1,\dots, \theta^K\}$ on
a grid around the mode so that they represent the density mass.  Then, we
approximate $\pi(\bm{\theta}|\bm{y})$ for each value
$\{\theta^1,\dots, \theta^K\}$.
In this special case the full conditional
$\pi(\bm{x}|\bm{y},\bm{\theta})$ is Gaussian and therefore
\eqref{eq:theta}  can be computed, for any value of $\theta$, without the need to approximate the
denominator.
If we are interested in the posterior marginal for $\theta$
we can
interpolate the points $\{\pi(\theta^1|\bm{y}),\dots, \pi(\theta^K|\bm{y})\}$  using, for example, a spline and normalize the
density.
Figure~\ref{Fig:ex1_2} shows the normalized density
$\pi(\theta|\bm{y})$. On the x-axis ten selected support points
$\theta^1, \ldots, \theta^{10}$ are indicated. The density line is obtained by fitting a
spline to $\{\theta^k,\log(\pi(\theta^k|y)\}$ and then normalizing so
that the density integrates to 1.

If we are interested in $\pi(x_i|\bm{y})$, the next task is to
approximate $\pi(x_i|\theta^k,\bm{y})$. Again, since the full
conditional of $\pi(\bm{x}|\theta^k,\bm{y})$ is Gaussian, its marginal
distributions can be easily found.
Finally we need to compute $\widetilde{\pi}(x_i|\bm{y})$ via
\eqref{eq.integrals2}.
Note that in this special case, the integrand can be computed exactly
for each value of $\theta$,
the only approximation error comes from the numerical integration
scheme in \eqref{eq:INLA_int}.  In the Gaussian likelihood case the
approximation $\widetilde{\pi}(x_i|\bm{y})$ is
a mixture of Gaussian densities $\pi(x_i|\bm{y}, \theta^k)$
weighted by $\tilde{\pi}(\theta^k|\bm{y})\Delta_k$, $k = 1,\dots,K$. Figure
\ref{Fig:ex1_3} shows the 10 elements of the mixture to
approximate $\pi(x_{10}|\bm{y})$ unweighted,  while Figure \ref{Fig:ex1_5} shows
the same elements but weighed. The sum of the densities in  Figure
\ref{Fig:ex1_5} gives the approximated posterior marginal
$\tilde{\pi}(x_{10}|\bm{y})$ also shown in Figure \ref{Fig:ex1_5}.
The procedure is repeated for each element of the vector
$\bm{x}$. These approximated densities can then be used to compute
posterior summary measures of interest. As an example, Figure \ref{Fig:ex1_1}
shows the posterior mean and 95\%
credible intervals for the underlying smooth function.

\section{Using the INLA framework in practice: The {\tt R-INLA} package}

The {\tt R-INLA} package provides a user friendly implementation of
the INLA methodology. 
It can be downloaded from \url{www.r-inla.org} together with the free-standing external INLA program.

The model definition in {\tt R-INLA} is similar to several other {\tt R}
packages, for example the {\tt mgcv} package to fit generalized
additive models  \citep{wood_glm}. There are two essential steps: 1)
Define the linear predictor through a {\tt
  formula} object; 2) Complete the model definition and fit the model using the function {\tt inla()}. The fitted model is returned as an {\tt inla} object.
The {\tt formula} can include fixed effects and
random effects.  Non-linear terms and random effects are included in the
formula using the {\tt f()} function. The specification of different
latent Gaussian models, hyperpriors and model fitting
options is straightforward. Results include the posterior marginal
distributions of the latent effects and hyperparameters, as well as
summary statistics. Furthermore, posterior estimates of linear combinations or
transformations of the latent field can be obtained \citep{newfeatures}.  Model choice
criteria such as the marginal likelihood,  DIC \citep{DIC2002}, WAIC \citep{Watanabe:2010},
conditional predictive ordinates and the probability integral
transform are also available. In {\tt R-INLA} there is  no function
``predict'' as for {\tt glm} or {\tt lm}. Predictions must be done as
a part of the model fitting itself. As prediction can be regarded as fitting a model with missing data,  we can simply set {\tt y[i] = NA} for those ``locations'' we want to predict. Predictive distributions, which are
often of interest, are however not returned directly. Instead the posterior marginals for random effects and the linear predictor at
the missing locations are returned. Adding the observational noise to
the fitted values leads to the
predictive distributions.

In the following we use a simple example of Poisson regression to illustrate the {\tt R-INLA} library.
In Section \ref{sec:prediction} we  illustrate how to
obtain the predictive distribution for a missing observation.

\subsection{Data preparation and model specification}

The {\tt Salm} \citep{Breslow1984}  dataset will be used throughout
this tutorial, the same dataset is  presented in both the  {\tt
  OpenBugs} tutorial and on {\tt R-INLA} example page on \url{www.r-inla.org}.
The data concern the   number of revertant colonies of TA98 Salmonella observed on each of three replicate plates tested at each of six dose levels of quinoline. A certain dose-response curve is suggested by theory and no effect of plate is assumed.

 Let  $y_{ij}$, $j=1,\dots,3$, $i = 1,\dots,6$ denote the number of colonies found on plate $j$ for dose $i$ and let $x_i$ indicate the  $i^{\text{th}}$ dose.
We assume a Poisson likelihood and use a random effect to allow for over-dispersion:
 \begin{align*}
y_{ij} \mid \lambda_{ij} &\sim\text{Poisson}(\lambda_{ij}) & i = 1,\dots,6\ j = 1,\dots,3\\
\log(\lambda_{ij})&  = \beta_0 + \beta_1 \log(x_{i}+10) + \beta_2 x_{i} + u_{ij}
 \end{align*}

 Here,   $\lambda_{ij}$ is the expected number of colonies at dose $i$
 for plate $j$. Further, $\beta_0$ denotes the intercept, $\beta_1$ and $\beta_2$
 fixed effects,  and
 $u_{ij}\mid \sigma^2 \sim \mathcal{N}(0, \sigma^2)$ is a random
 effect accounting for unobserved heterogeneity.
 Putting this model in the LGMs framework described in Section
 \ref{Section:LGM}, the latent Gaussian field is $\{\lambda_{11},\dots,\lambda_{36},\beta_0,
 \beta_1,\beta_2,u_{11},\dots,u_{36}\}$. The model has one hyperparameter,
 namely the variance $\sigma^2$  of the random effect $\bm{u}$. To
 complete the model we need to define a prior distribution for
 $\sigma^2$. In the INLA world priors are usually not defined on
 variances but on its inverse, the precision parameters: here we
 use a PC-prior \citep{simpson2017} with parameters $u=1$ and $\alpha=0.01$ as prior for the log precision $\log(\tau) = -2\log(\sigma)$.

The model is specified as follows as a {\tt formula} object:
\begin{knitrout}
\definecolor{shadecolor}{rgb}{0.969, 0.969, 0.969}\color{fgcolor}\begin{kframe}
\begin{alltt}
\hlcom{# load the data set}
\hlkwd{data}\hlstd{(Salm)}
\hlcom{# rename the columns to fit the notation}
\hlkwd{names}\hlstd{(Salm)} \hlkwb{=} \hlkwd{c}\hlstd{(}\hlstr{"y"}\hlstd{,} \hlstr{"x"}\hlstd{,} \hlstr{"u"}\hlstd{)}
\hlkwd{head}\hlstd{(Salm)}
\end{alltt}
\begin{verbatim}
##    y  x u
## 1 15  0 1
## 2 21  0 2
## 3 29  0 3
## 4 16 10 4
## 5 18 10 5
## 6 21 10 6
\end{verbatim}
\begin{alltt}
\hlcom{# specify the prior for the log precision parameter}
\hlstd{my.hyper} \hlkwb{<-} \hlkwd{list}\hlstd{(}\hlkwc{theta} \hlstd{=} \hlkwd{list}\hlstd{(}\hlkwc{prior}\hlstd{=}\hlstr{"pc.prec"}\hlstd{,} \hlkwc{param}\hlstd{=}\hlkwd{c}\hlstd{(}\hlnum{1}\hlstd{,}\hlnum{0.01}\hlstd{)))}
\hlcom{# specify the linear predictor}
\hlstd{formula} \hlkwb{<-} \hlstd{y} \hlopt{~} \hlkwd{log}\hlstd{(x} \hlopt{+} \hlnum{10}\hlstd{)} \hlopt{+} \hlstd{x} \hlopt{+} \hlkwd{f}\hlstd{(u,} \hlkwc{model} \hlstd{=} \hlstr{"iid"}\hlstd{,} \hlkwc{hyper} \hlstd{= my.hyper)}
\end{alltt}
\end{kframe}
\end{knitrout}
Of note, an intercept is automatically included and can be removed by
adding ``-1'' or ``0'' to the {\tt formula}.
The function {\tt inla.list.models()} provides a list of available
distributions for the different parts of the model. Possible parameters for the {\tt inla.list.models()}  function are {\tt "prior"} (available priors for the hyperparameters), {\tt "likelihood"} (all implemented likelihoods) and {\tt "latent"} (available models for the latent field).

The {\tt f()} function is used to specify the latent Gaussian model
for the random effect, here an independent noise model, and the
hyperprior for its corresponding hyperparameters.
Information about the different latent Gaussian  models can be obtained through the function {\tt inla.doc()}, for example:
\begin{knitrout}
\definecolor{shadecolor}{rgb}{0.969, 0.969, 0.969}\color{fgcolor}\begin{kframe}
\begin{alltt}
\hlkwd{inla.doc}\hlstd{(}\hlstr{"iid"}\hlstd{)}
\end{alltt}
\end{kframe}
\end{knitrout}
will provide information about the {\tt iid} model.

 The formula object is further fed to the main function {\tt inla()}:

\begin{knitrout}
\definecolor{shadecolor}{rgb}{0.969, 0.969, 0.969}\color{fgcolor}\begin{kframe}
\begin{alltt}
\hlstd{result} \hlkwb{<-} \hlkwd{inla}\hlstd{(}\hlkwc{formula}\hlstd{=formula,} \hlkwc{data}\hlstd{=Salm,} \hlkwc{family}\hlstd{=}\hlstr{"Poisson"}\hlstd{)}
\end{alltt}
\end{kframe}
\end{knitrout}

It requires as first argument the formula object. Furthermore,
 the likelihood must be specified in form of a string and the data object
 must be specified which needs to be a {\tt data.frame} or {\tt list}. The variable
 names used in the data object must of course fit the notation used in
 the formula object. Within the {\tt inla} function different {\tt
   control.*} statements can be included. Examples are {\tt
   control.compute = list(dic=TRUE, waic=TRUE}) to obtain DIC and WAIC
 measures, {\tt control.inla=list(int.strategy='eb')} to change  to
 the empirical Bayes strategy when doing the integration over the
 hyperparameter space, or {\tt control.fixed=list(\ldots)} to
 change the default prior specification for the fixed effects. Within
 {\tt R} documentation is provided by typing {\tt ?control.fixed} for
 example.

\subsection{Getting Results}
The {\tt R-INLA} object {\tt result} contains all results. First,
we can look at some posterior summary information using
\begin{knitrout}
\definecolor{shadecolor}{rgb}{0.969, 0.969, 0.969}\color{fgcolor}\begin{kframe}
\begin{alltt}
\hlkwd{summary}\hlstd{(result)}
\end{alltt}
\begin{verbatim}
## 
## Call:
## "inla(formula = formula, family = \"Poisson\", data = Salm)"
## 
## Time used:
##  Pre-processing    Running inla Post-processing           Total 
##           1.150           0.092           0.065           1.307 
## 
## Fixed effects:
##               mean     sd 0.025quant 0.5quant 0.97quant   mode kld
## (Intercept)  2.168 0.3588     1.4507    2.170    2.8432  2.174   0
## log(x + 10)  0.313 0.0976     0.1188    0.313    0.4980  0.313   0
## x           -0.001 0.0004    -0.0018   -0.001   -0.0002 -0.001   0
## 
## Random effects:
## Name	  Model
##  u   IID model 
## 
## Model hyperparameters:
##                  mean    sd 0.025quant 0.5quant 0.97quant  mode
## Precision for u 20.84 18.25      5.718    16.46     57.56 11.92
## 
## Expected number of effective parameters(std dev): 12.05(2.081)
## Number of equivalent replicates : 1.494 
## 
## Marginal log-Likelihood:  -83.68
\end{verbatim}
\end{kframe}
\end{knitrout}
This provides information about the processing time and some
statistics about the posterior distributions of the fixed effects and
the hyperparameter. The random effects are only listed by name together with
their prior model. Posterior marginals for the fixed effects, random
effects, hyperparameters, and so on, can be
found in  {\tt result\$marginals.fixed}, {\tt
  result\$marginals.random},
{\tt result\$marginals.hyperpar}, respectively, while posterior summary information
is provided in {\tt result\$summary.fixed}, {\tt
  result\$summary.random},
{\tt result\$summary.hyperpar}, respectively.
Note, by default
 INLA provides posterior summary information for precision parameters,
 i.e. inverse variance parameters. However, using functions such
 as {\tt inla.emarginal()} and {\tt inla.tmarginal()} posterior information on
 standard deviation or variance scale can be easily obtained.
  The posterior marginal for $\tau = 1/\sigma^2$ is saved in {\tt result\$marginals.hyperpar}, which is a list of length equal to the number of model hyperparameters. The following chunk of code illustrates how to get the posterior mean and standard deviation for the standard deviation $\sigma$:
\begin{knitrout}
\definecolor{shadecolor}{rgb}{0.969, 0.969, 0.969}\color{fgcolor}\begin{kframe}
\begin{alltt}
\hlcom{# Select the right hyperparameter marginal}
\hlstd{tau} \hlkwb{<-} \hlstd{result}\hlopt{$}\hlstd{marginals.hyperpar[[}\hlnum{1}\hlstd{]]}
\hlcom{# Compute the expected value for 1/\textbackslash{}sqrt\{\textbackslash{}tau\} and 1/\textbackslash{}sqrt\{tau\}^2}
\hlstd{E} \hlkwb{=} \hlkwd{inla.emarginal}\hlstd{(}\hlkwa{function}\hlstd{(}\hlkwc{x}\hlstd{)} \hlkwd{c}\hlstd{(}\hlnum{1}\hlopt{/}\hlkwd{sqrt}\hlstd{(x),(}\hlnum{1}\hlopt{/}\hlkwd{sqrt}\hlstd{(x))}\hlopt{^}\hlnum{2}\hlstd{), tau)}
\hlcom{# From this we computed the posterior standard deviation as}
\hlstd{mysd} \hlkwb{=} \hlkwd{sqrt}\hlstd{(E[}\hlnum{2}\hlstd{]} \hlopt{-} \hlstd{E[}\hlnum{1}\hlstd{]}\hlopt{^}\hlnum{2}\hlstd{)}
\hlcom{# so that we obtain the posterior mean and standard deviation}
\hlkwd{print}\hlstd{(}\hlkwd{c}\hlstd{(}\hlkwc{mean}\hlstd{=E[}\hlnum{1}\hlstd{],} \hlkwc{sd}\hlstd{=mysd))}
\end{alltt}
\begin{verbatim}
##  mean    sd 
## 0.253 0.074
\end{verbatim}
\end{kframe}
\end{knitrout}
If we were interested not only in some summary statistics but in the whole posterior density of the standard deviation, we can use the {\tt inla.tmarginal()} function as follows:
\begin{knitrout}
\definecolor{shadecolor}{rgb}{0.969, 0.969, 0.969}\color{fgcolor}\begin{kframe}
\begin{alltt}
\hlstd{my.sigma} \hlkwb{<-} \hlkwd{inla.tmarginal}\hlstd{(}\hlkwa{function}\hlstd{(}\hlkwc{x}\hlstd{)\{}\hlnum{1}\hlopt{/}\hlkwd{sqrt}\hlstd{(x)\}, tau)}
\end{alltt}
\end{kframe}
\end{knitrout}
Figure \ref{fig:hyperpar} shows the posterior marginals
$\pi(\tau|\bm{y})$ and $\pi(\sigma|\bm{y})$ as computed above.
\begin{knitrout}
\definecolor{shadecolor}{rgb}{0.969, 0.969, 0.969}\color{fgcolor}\begin{figure}

{\centering \includegraphics[width=.48\linewidth]{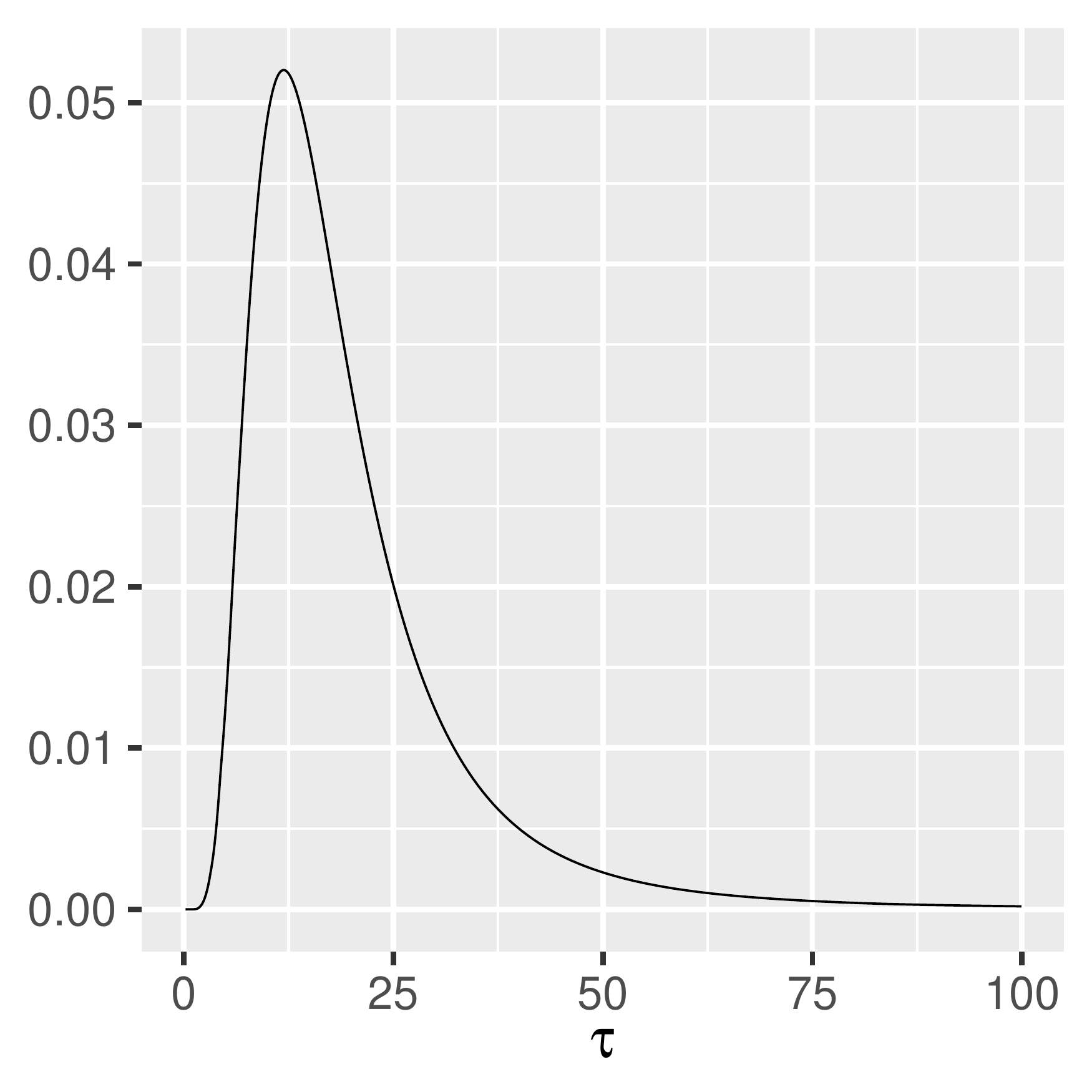} 
\includegraphics[width=.48\linewidth]{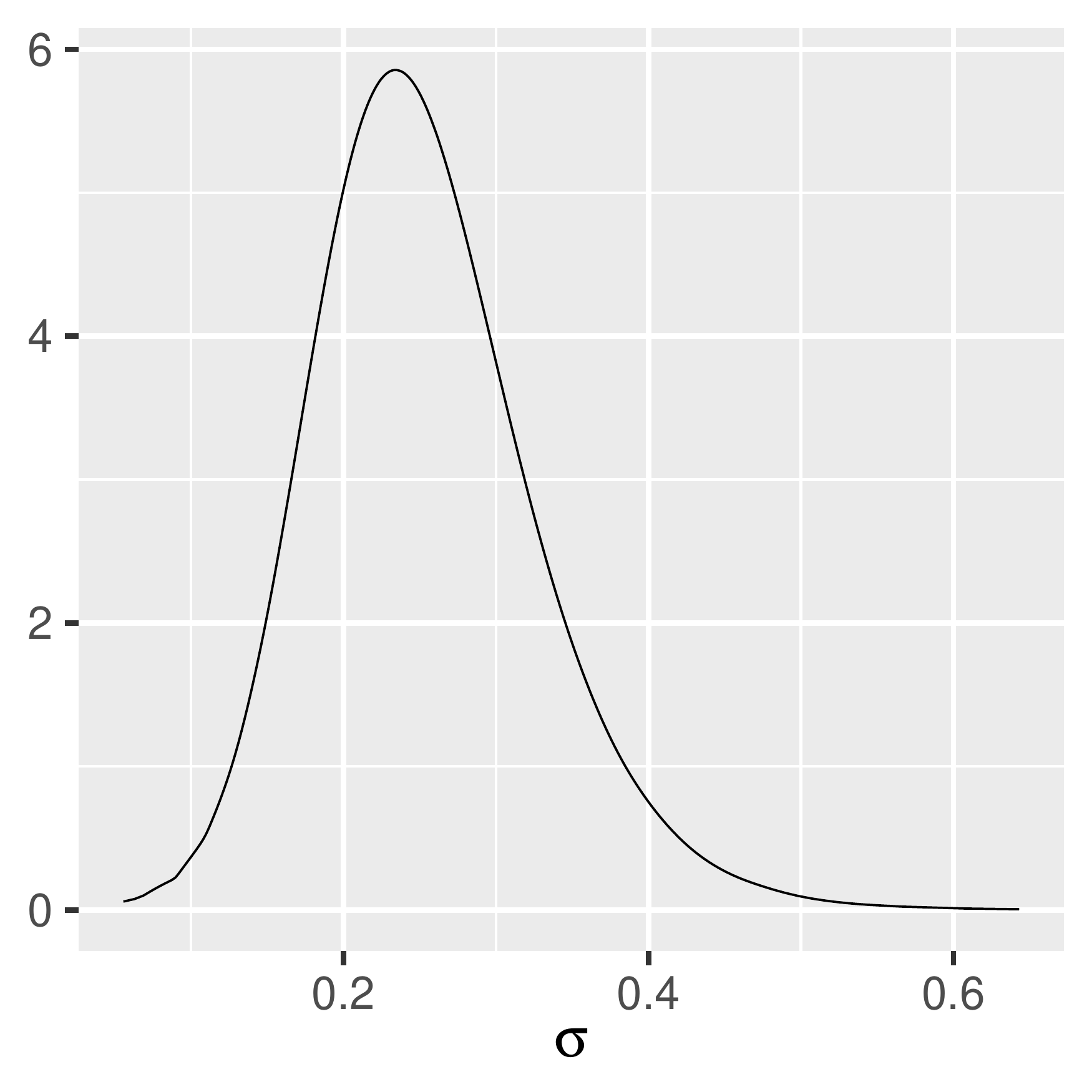} 

}

\caption[Posterior marginal for the hyperparameter]{Posterior marginal for the hyperparameter: on the precision scale (left) and on the standard deviation scale (right).}\label{fig:hyperpar}
\end{figure}

\end{knitrout}
 From the transformed posterior marginal the function
 {\tt inla.zmarginal()} allows  to extract posterior 'z'ummary information:
\begin{knitrout}
\definecolor{shadecolor}{rgb}{0.969, 0.969, 0.969}\color{fgcolor}\begin{kframe}
\begin{alltt}
\hlkwd{inla.zmarginal}\hlstd{(my.sigma)}
\end{alltt}
\begin{verbatim}
## Mean            0.253194 
## Stdev           0.0735528 
## Quantile  0.025 0.127062 
## Quantile  0.25  0.202214 
## Quantile  0.5   0.246286 
## Quantile  0.75  0.296463 
## Quantile  0.975 0.417444
\end{verbatim}
\end{kframe}
\end{knitrout}

The {\tt inla.emarginal()} and {\tt inla.tmarginal()} functions we used above are part of a family of {\tt inla.$\star$marginal} functions in the {\tt R-INLA} library that can be used to manipulate univariate posterior marginals in different ways. Table \ref{tab:margFunc} provides a list of such functions with the relative usage.
\begin{table}[h]
  \begin{center}
    \begin{tabular}{l|p{8cm}}
      \bf{Function Name} & \bf{Usage}\\\hline\hline
      \tt{inla.dmarginal(x, marginal, $\dots$)} & Density at a vector of
      evaluation points $x$ \\
      \tt{inla.pmarginal(q, marginal, $\dots$)} & Distribution function at a vector of
      quantiles $q$ \\
      \tt{inla.qmarginal(p, marginal, $\dots$)} & Quantile function at a vector of
      probabilities $p$.\\
      \tt{inla.rmarginal(n, marginal)} & Generate $n$ random deviates \\
      \tt{inla.hpdmarginal(p, marginal, $\dots$)} & Compute the highest posterior density
      interval at level $p$\\
      \tt{inla.emarginal(fun, marginal, $\dots$)} & Compute the expected value of the
      marginal assuming the transformation given by fun\\
      \tt{inla.mmarginal(marginal)} & Computes the mode\\
      \tt{inla.smarginal(marginal, $\dots$)} & Smoothed density in
      form of a list of length two. The first entry contains the x-values, the second
      entry includes the interpolated y-values\\
      \tt{inla.tmarginal(fun, marginal, $\dots$)} & Transform the marginal using the
      function fun.\\
      \tt{inla.zmarginal(marginal)} & Summary statistics for the marginal\\
      \bottomrule
    \end{tabular}
    \caption{Functions which use a posterior marginal density to derive some
    information of interest. The {\tt marginal} is thereby given
    in form of a matrix with two columns where the first column
    represents the location points and the second column the density
    values at those location points. \label{tab:margFunc}}
  \end{center}
\end{table}

Posterior marginals for the fixed effects are stored in {\tt result\$marginals.fixed}. This is a list of length equal to the number of fixed effects in the model plus the intercept (3 in this case). We can obtain the posterior mean of the intercept as follows:
\begin{knitrout}
\definecolor{shadecolor}{rgb}{0.969, 0.969, 0.969}\color{fgcolor}\begin{kframe}
\begin{alltt}
\hlkwd{inla.emarginal}\hlstd{(}\hlkwa{function}\hlstd{(}\hlkwc{x}\hlstd{) x, result}\hlopt{$}\hlstd{marginals.fixed}\hlopt{$}\hlstd{`(Intercept)`)}
\end{alltt}
\begin{verbatim}
## [1] 2.2
\end{verbatim}
\end{kframe}
\end{knitrout}
However, note that summary information for all the fixed effects is
directly available in the slot {\tt result\$summary.fixed}.
\begin{knitrout}
\definecolor{shadecolor}{rgb}{0.969, 0.969, 0.969}\color{fgcolor}\begin{kframe}
\begin{alltt}
\hlstd{result}\hlopt{$}\hlstd{summary.fixed}
\end{alltt}
\begin{verbatim}
##                 mean      sd 0.025quant 0.5quant 0.97quant     mode
## (Intercept)  2.16813 0.35883     1.4507  2.17009   2.84317  2.17401
## log(x + 10)  0.31294 0.09764     0.1188  0.31300   0.49800  0.31313
## x           -0.00098 0.00043    -0.0018 -0.00098  -0.00016 -0.00098
##                 kld
## (Intercept) 7.6e-07
## log(x + 10) 2.1e-06
## x           2.1e-06
\end{verbatim}
\end{kframe}
\end{knitrout}

 Finally, the marginal densities for the random effects are stored in
 {\tt result\$marginals.random\$plate} (a list with length $18$ elements in this case) while summary statistics for the random effects are stored in {\tt result\$summary.random\$rand}.
 Figure \ref{fig:random} shows the posterior mean of $\bm{u}$ within $2.5\%$ and $97.5\%$ quantiles  and is created  from information stored in the
 {\tt result\$summary.random\$u} object.
\begin{knitrout}
\definecolor{shadecolor}{rgb}{0.969, 0.969, 0.969}\color{fgcolor}\begin{figure}

{\centering \includegraphics[width=.48\linewidth]{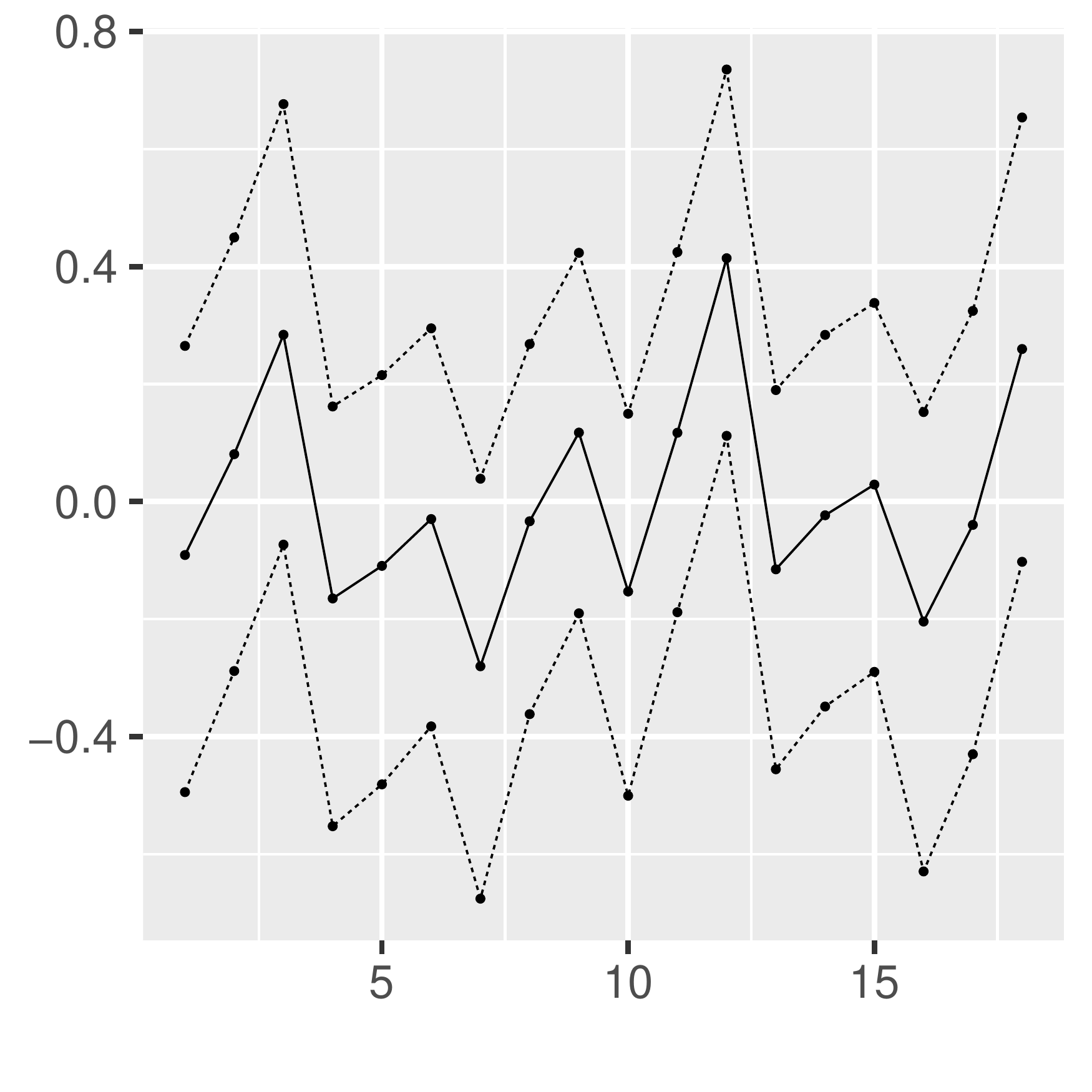} 

}

\caption[Posterior mean (solid line) together with  $2.5\%$ and $97.5\%$ posterior quantiles (broken lines) for the random effect $\bf{u}$]{Posterior mean (solid line) together with  $2.5\%$ and $97.5\%$ posterior quantiles (broken lines) for the random effect $\bf{u}$}\label{fig:random}
\end{figure}

\end{knitrout}

\subsection{Getting prediction densities}\label{sec:prediction}

Posterior predictive distributions are of interest in many applied
problems. The {\tt inla()} function does not return predictive
densities. In this Section we show how to post-process the result from
{\tt inla()} in order to obtain the posterior predictive density for a
point of interest.

Assume we would like to predict a new datapoint. For ease of
illustration we remove datapoint $k = 7$ in the {\tt Salm} dataset and predict it. That is, we are interested in $\pi(y_7|\bm{y}_{-7})$ where $\bm{y}_{-7}$ is the vector of all observations except for the $7^{\text{th}}$.

The {\tt inla()} function allows for missing values in the response
variable, and  computes the posterior marginal for the corresponding linear predictor. We create a new dataset {\tt Salm.predict} where observation 7 is set to {\tt NA} and rerun the {\tt inla()} function:
\begin{knitrout}
\definecolor{shadecolor}{rgb}{0.969, 0.969, 0.969}\color{fgcolor}\begin{kframe}
\begin{alltt}
\hlcom{## set observation 7 to NA}
\hlstd{Salm.predict} \hlkwb{=} \hlstd{Salm}
\hlstd{Salm.predict[}\hlnum{7}\hlstd{,} \hlstr{"y"}\hlstd{]} \hlkwb{<-} \hlnum{NA}
\hlcom{# re-run the model}
\hlstd{res.predict} \hlkwb{=} \hlkwd{inla}\hlstd{(}\hlkwc{formula}\hlstd{=formula,} \hlkwc{data}\hlstd{=Salm,}
  \hlkwc{family}\hlstd{=}\hlstr{"Poisson"}\hlstd{,}
  \hlkwc{control.predictor} \hlstd{=} \hlkwd{list}\hlstd{(}\hlkwc{compute} \hlstd{=} \hlnum{TRUE}\hlstd{),}
  \hlkwc{control.family} \hlstd{=} \hlkwd{list}\hlstd{(}\hlkwc{control.link}\hlstd{=}\hlkwd{list}\hlstd{(}\hlkwc{model}\hlstd{=}\hlstr{"log"}\hlstd{))}
  \hlstd{)}
\end{alltt}
\end{kframe}
\end{knitrout}
Note that, compared to the previous run of {\tt inla()}, here we have
specified some extra parameters. As default, the  posterior marginals
for the linear predictor are not provided. By specifying {\tt
  control.predictor = list(compute = TRUE)} the posterior marginals
will be included in the results object. We also need to explicitly
specify the link function using the {\tt control.family} object in
order for {\tt inla()} to compute the linear predictor not only at the
linear scale ($\eta$) but also at the observations scale ($\lambda = \exp(\eta)$).

We can inspect the linear predictor $\eta_7 = \log(\lambda_7)$ using
\begin{knitrout}
\definecolor{shadecolor}{rgb}{0.969, 0.969, 0.969}\color{fgcolor}\begin{kframe}
\begin{alltt}
\hlcom{# marginal posterior for the linear predictor}
\hlstd{eta7} \hlkwb{=} \hlstd{res.predict}\hlopt{$}\hlstd{marginals.linear.predictor[[}\hlnum{7}\hlstd{]]}
\hlcom{# some summary statistics}
\hlkwd{round}\hlstd{(res.predict}\hlopt{$}\hlstd{summary.linear.predictor[}\hlnum{7}\hlstd{,],} \hlnum{3}\hlstd{)}
\end{alltt}
\begin{verbatim}
##              mean   sd 0.025quant 0.5quant 0.97quant mode kld
## Predictor.07    3 0.18        2.6        3       3.4  3.1   0
\end{verbatim}
\end{kframe}
\end{knitrout}
We can then compute the linear predictor at the observation scale $\lambda_7 = \exp(\eta_7)$ using the {\tt inla.tmarginal()} function as before. An alternative, having specified the link function in the {\tt control.family()} parameter, is to extract $\lambda_7$ directly from the result object:
\begin{knitrout}
\definecolor{shadecolor}{rgb}{0.969, 0.969, 0.969}\color{fgcolor}\begin{kframe}
\begin{alltt}
\hlcom{# extract from the res.predict object}
\hlstd{lambda7} \hlkwb{=} \hlstd{res.predict}\hlopt{$}\hlstd{marginals.fitted.values[[}\hlnum{7}\hlstd{]]}
\hlcom{# compute using inla.tmarginal()}
\hlstd{lambda7_bis} \hlkwb{=} \hlkwd{inla.tmarginal}\hlstd{(}\hlkwa{function}\hlstd{(}\hlkwc{x}\hlstd{)\{}\hlkwd{exp}\hlstd{(x)\},}
  \hlstd{res.predict}\hlopt{$}\hlstd{marginals.linear.predictor[[}\hlnum{7}\hlstd{]])}
\end{alltt}
\end{kframe}
\end{knitrout}
Figure \ref{fig:linearPredictor} shows the posterior marginal
$\widetilde{\pi}(\eta_7|\bm{y})$ and
$\widetilde{\pi}(\lambda_7|\bm{y})$.
\begin{knitrout}
\definecolor{shadecolor}{rgb}{0.969, 0.969, 0.969}\color{fgcolor}\begin{figure}

{\centering \includegraphics[width=.48\linewidth]{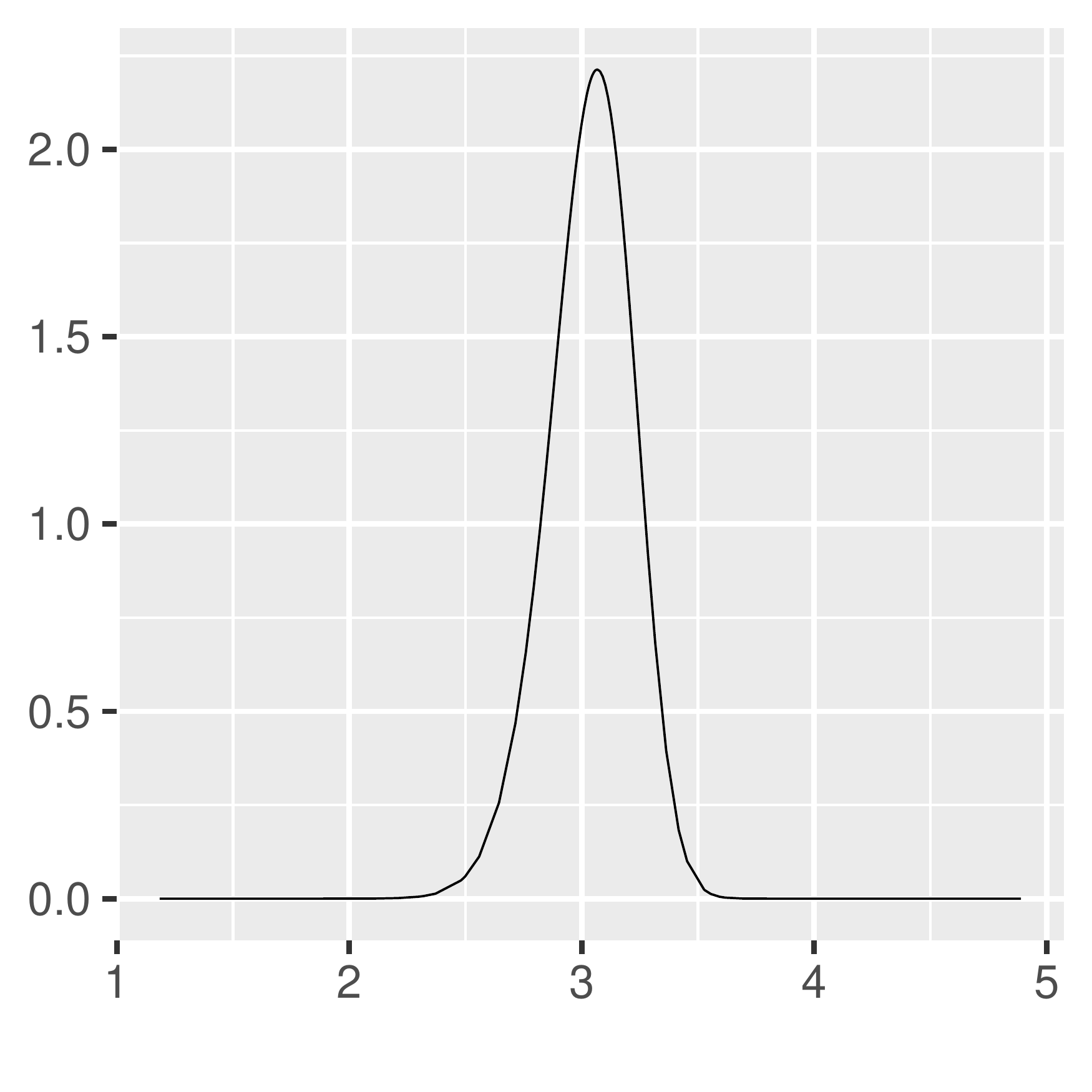} 
\includegraphics[width=.48\linewidth]{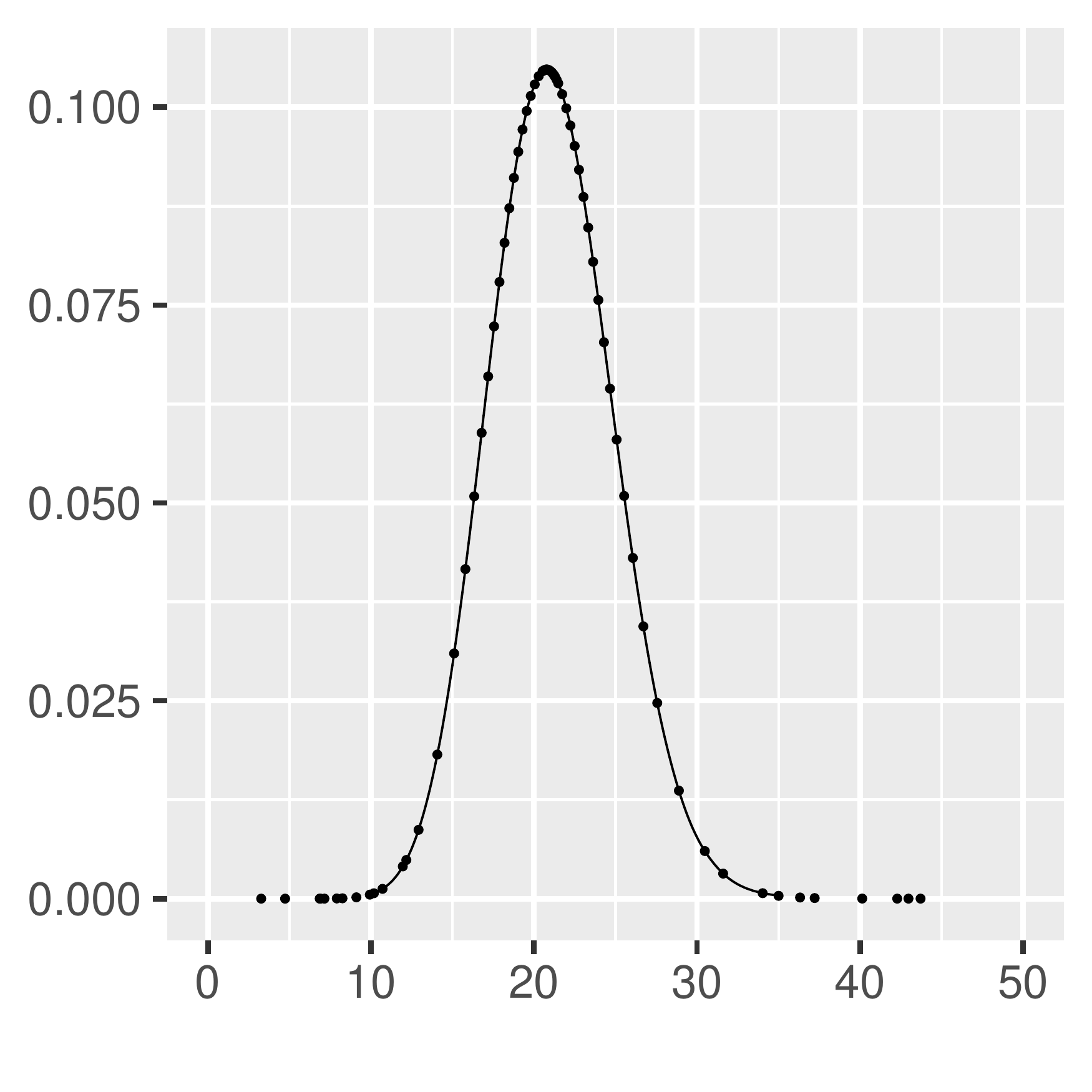} 

}

\caption{Posterior marginal for the linear predictor (left) and for the linear predictor at the observation scale, i.e.~the fitted value scale (right). The last quantity can be directly extracted from the result object (dots) or computed using the {\tt inla.tmarginal()} function (black line).}\label{fig:linearPredictor}
\end{figure}

\end{knitrout}
The predictive distribution is given by:
\begin{equation}\label{eq:predictive}
  \widetilde{\pi}(y_7|\bm{y}_{-7}) = \int \widetilde{\pi}(y_7|\lambda_{7})  \widetilde{\pi}(\lambda_7|\bm{y}_{-7}) d\lambda_7
\end{equation}
We can solve \eqref{eq:predictive} in two ways, by sampling or numerical integration.
If we want to sample we need to proceed in two steps: first sample values for $\lambda_7$ from its posterior density, then sample possible observations from a Poisson likelihood with mean equal to the sampled values of $\lambda_7$:
\begin{knitrout}
\definecolor{shadecolor}{rgb}{0.969, 0.969, 0.969}\color{fgcolor}\begin{kframe}
\begin{alltt}
\hlstd{n.samples} \hlkwb{=} \hlnum{3000}
\hlstd{samples_lambda} \hlkwb{=} \hlkwd{inla.rmarginal}\hlstd{(n.samples, lambda7)}
\hlcom{# sample from the likelihood model}
\hlstd{predDist} \hlkwb{=} \hlkwd{rpois}\hlstd{(n.samples,} \hlkwc{lambda} \hlstd{= samples_lambda)}
\end{alltt}
\end{kframe}
\end{knitrout}
Figure \ref{fig:prediction} shows an histogram of the predicted values sampled as above.
Note that, in this case, we could sample directly from the posterior
marginal of the linear predictor at the observation scale. When the
likelihood depends not only in the latent field but also on some
hyperparameters (as in the Gaussian case, say) we need to sample both $\eta$ and $\bm{\theta}$ jointly. This can be done using the function
{\tt inla.posterior.sample()} which takes as input the number of samples and the {\tt res.predict} object and return samples from the joint posterior marginal. In order for this function to work one has to provide the extra parameter {\tt control.compute = list(config = TRUE)} when calling the {\tt inla()} function.

A second possibility is to solve \eqref{eq:predictive} numerically as:
\begin{knitrout}
\definecolor{shadecolor}{rgb}{0.969, 0.969, 0.969}\color{fgcolor}\begin{kframe}
\begin{alltt}
 \hlcom{# library supporting trapezoid rule integration.}
 \hlkwd{library}\hlstd{(caTools)}

 \hlcom{# specify the support at which we want to compute the density}
 \hlstd{ii} \hlkwb{=} \hlnum{0}\hlopt{:}\hlnum{100}
 \hlstd{predDist2} \hlkwb{=} \hlkwd{rep}\hlstd{(}\hlnum{0}\hlstd{,}\hlnum{101}\hlstd{)}
 \hlcom{# go over the posterior marginal of the fitted value}
\hlkwa{for}\hlstd{(j} \hlkwa{in} \hlnum{1}\hlopt{:}\hlstd{(}\hlkwd{length}\hlstd{(lambda7[ ,}\hlnum{1}\hlstd{])}\hlopt{-}\hlnum{1}\hlstd{))\{}
  \hlstd{predDist2} \hlkwb{<-} \hlstd{predDist2} \hlopt{+} \hlkwd{dpois}\hlstd{(ii,}
           \hlkwc{lambda} \hlstd{= ((lambda7[j,}\hlnum{1}\hlstd{]}\hlopt{+} \hlstd{lambda7[j}\hlopt{+}\hlnum{1}\hlstd{,}\hlnum{1}\hlstd{])}\hlopt{/}\hlnum{2}\hlstd{))} \hlopt{*}
             \hlkwd{trapz}\hlstd{(lambda7[j}\hlopt{:}\hlstd{(j}\hlopt{+}\hlnum{1}\hlstd{),} \hlnum{1}\hlstd{], lambda7[j}\hlopt{:}\hlstd{(j}\hlopt{+}\hlnum{1}\hlstd{),} \hlnum{2}\hlstd{])}
 \hlstd{\}}
\end{alltt}
\end{kframe}
\end{knitrout}
A drawback with this method is that one has to locate the area of high
density of the predictive distribution in order to perform the
integration. 
Figure \ref{fig:prediction} shows the prediction distribution as computed above.

\begin{knitrout}
\definecolor{shadecolor}{rgb}{0.969, 0.969, 0.969}\color{fgcolor}\begin{figure}

{\centering \includegraphics[width=.48\linewidth]{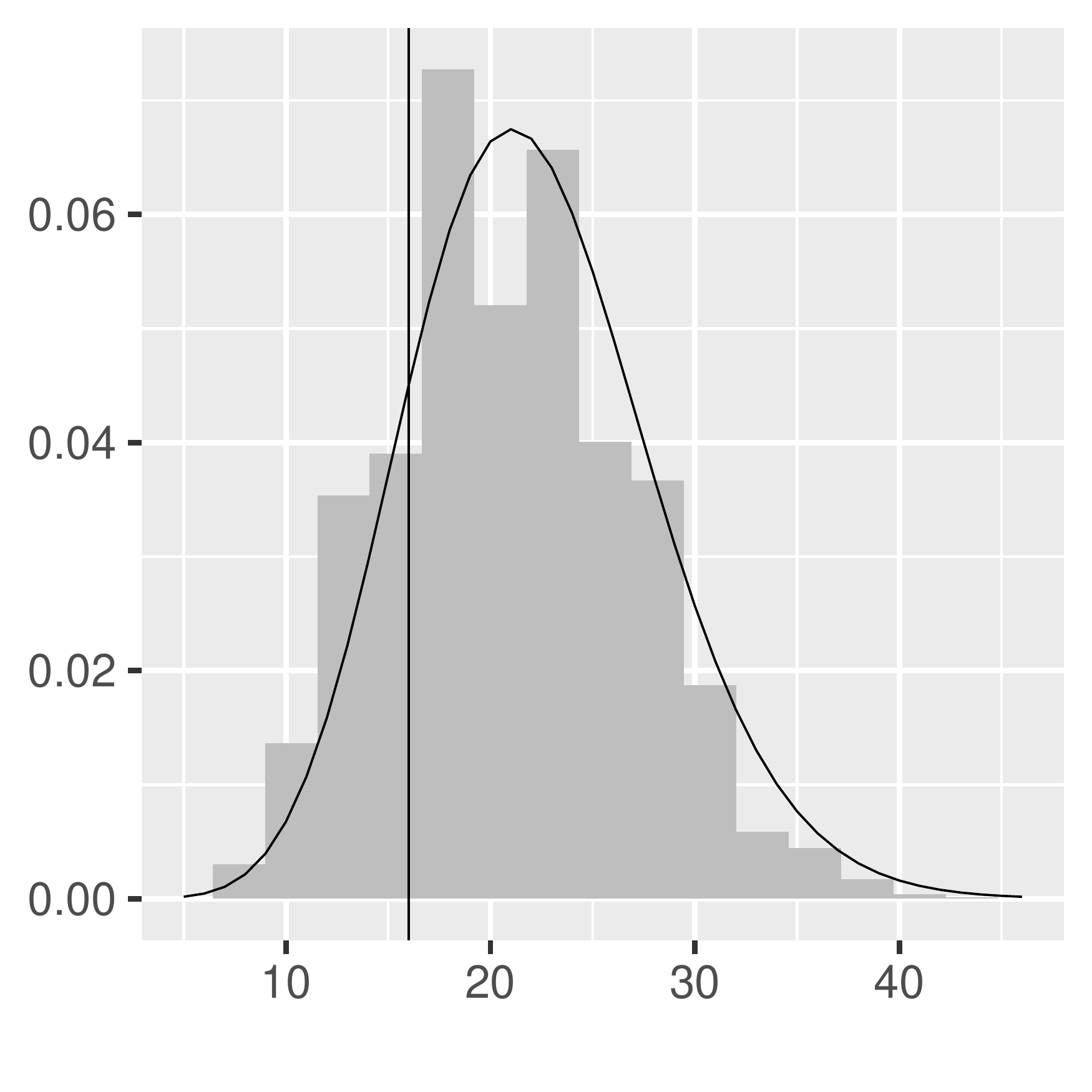} 

}

\caption[Estimate of the posterior predictive density  $\widetilde{\pi}(y_7|\bm{y}_{-7})$ via sampling (histogram) and numerical integration (black line)]{Estimate of the posterior predictive density  $\widetilde{\pi}(y_7|\bm{y}_{-7})$ via sampling (histogram) and numerical integration (black line). The vertical line indicate the observation $y_7$ that we have removed from the dataset.}\label{fig:prediction}
\end{figure}

\end{knitrout}

\section{Expanding the scope: INLA within MCMC}

Implementing INLA from scratch is a complex task so that, in practice,
the applications of INLA are limited to the (large class of) models
implemented in the {\tt R-INLA} library. The library allows the user
to manually specify latent Gaussian models, see {\tt
  vignette("rgeneric", package="INLA")}, that are not directly
available within the library.
Also user-defined hyperpriors are possible. However, certain models do
not fit into the scope of {\tt R-INLA}. Consider for example the case
in which  one observation does
not only depend on exactly one element of the linear predictor.

 Recently \citet{Gomez-Rubio2018} proposed a novel approach that
 enlarges the class of models that can  benefit from the fast computations of INLA.
  Let $\bm{z} = (\bm{x},\bm{\theta})$ denote all model parameters. \citet{BIVAND2014146}, noticed that some models can be fit with {\tt R-INLA} after conditioning on one or several parameters. That is, we split the parameter vector into two, $\bm{z} = (\bm{z}_c, \bm{z}_{-c})$, and assume that the model
\[
  \pi(\bm{z}_{-c}|\bm{y},\bm{z}_c)\propto \pi(\bm{y}|\bm{z}_{-c}, \bm{z}_{c}) \pi(\bm{z}_{-c}| \bm{z}_{c})
\]
can be fit with {\tt R-INLA}.
If the conditioning parameters are few and bounded, one way to solve the full model is to define a grid  $\bm{z}^k_{c}$, $k = 1,\dots,K$ on the bounded domain of the conditioning parameters and run {\tt inla()} on
all $K$ conditional models. At each run we  get approximations to
the marginal likelihood $\tilde{\pi}(\bm{y}|\bm{z}_{c}^k)$ and to the
(conditional) posterior marginals
$\tilde{\pi}(z_{-c,j}|\bm{y},\bm{z}_c^k)$ for each element $j$ of vector $\bm{z}_{-c}$.  The posterior  for the conditioning parameters $\bm{z}_c$ can be obtained combining the marginal likelihood with the prior:
\[
\tilde{\pi}(\bm{z}_c^k|\bm{y})\propto\tilde{\pi}(\bm{y}|\bm{z}_{c}^k)\pi(\bm{z}_{c}^k)
\]
This can be easily normalized since we assume a bounded domain. For
the remainder of the parameters $\bm{z}_{-c}$, their posterior
marginal distribution can be obtained by Bayesian model averaging the
family of models fitted with {\tt R-INLA}:
 \begin{equation}\label{eq:BMA}
 \pi(z_{-c,j}|\bm{y}) = \int \pi(z_{-c,j}|\bm{y},\bm{z}_c) \pi(\bm{z}_c|\bm{y}) d\bm{z}_c \approx \sum_k \tilde{\pi}(z_{-c,j}|\bm{y},\bm{z}_c^{k}) \tilde{\pi}(\bm{z}_c^k|\bm{y}).
 \end{equation}
 This  solution,  becomes infeasible if  $\bm{z}_{c}$ is larger and 
unbounded.

 \citet{Gomez-Rubio2018}  propose instead to embed INLA in a larger
 Metropolis–Hastings (MH) algorithm \citep{doi:10.1063/1.1699114,
   hastings}.  In this way, the posterior marginals of an
 ensemble of parameters (namely those we condition on) can be obtained via MCMC sampling, whereas the posterior marginals of all the other parameters are obtained by averaging over several conditional marginal distributions.

To estimate the posterior distribution of all parameters in the model, the  MH algorithm can be used to draw values  for $\bm{z}_c$. At step $i$, new values $\bm{z}_c^{(i)}$ are proposed and accepted with probability
\begin{equation}\label{eq:proposal}
 \alpha = \min\left\{1, \frac{\pi(\bm{y}|\bm{z}_c^{(i)}) \pi(\bm{z}_c^{(i)})}{\pi(\bm{y}|\bm{z}_c^{(i-1)}) \pi(\bm{z}_c^{(i-1)})} \frac{q(\bm{z}_c^{(i-1)}|\bm{z}_c^{(i)})}{q(\bm{z}_c^{(i)}|\bm{z}_c^{(i-1)})}
  \right\}
\end{equation}
 If the proposal is not accepted, then $\bm{z}_c^{(i)}$ is set to  $\bm{z}_c ^{(i-1)}$.

 For each proposed value of the conditioning parameters $\bm{z}_c^{(i)}$ we run {\tt inla()} on the conditional model and obtain the approximated posterior (conditional) marginals for all parameters in $\bm{z}_{-c}$,  $\tilde{\pi}(z_{-c,j}|\bm{y},\bm{z}_c^{(i)})$. Note that, in this context we are free to choose any prior for $\bm{z}_c$.

After a suitable number of iterations, the MH algorithm will produce $N$ samples from $\pi(\bm{z}_c|\bm{y})$ which can be used to derive posterior marginals or any other summary statistics of interest. The posterior marginals for the parameters in $\bm{z}_{-c}$ can be approximated as
\[
\tilde{\pi}(z_{-c,j}|\bm{y}) = \frac{1}{N}\sum_i \tilde{\pi}(z_{-c,j}|\bm{y},\bm{z}_c^{(i)}) .
\]
 Note that using INLA within MCMC allows to perform multivariate posterior inference on the parameter subset $\bm{z}_c$.

 \citet{Gomez-Rubio2018}  illustrate the use of INLA within MCMC with
 different examples including some spatial econometrics models,
 Bayesian lasso and imputation of missing covariates. They show that
 the new algorithm provides accurate approximations to the posterior
 distribution. \citet{wakefield2019} use INLA within MCMC to correct
 for jittering in the locations of complex survey studies.
 INLA within MCMC can be used to fit models where non-Gaussian or
 multivariate priors are used on some elements of the latent field and
 hyperparameters. It can also be used in cases where the user wants to model both parameters in the likelihood (for example the mean and the standard deviation in the Gaussian likelihood) with respect to some covariates.
 This method largely increases the class of models that can benefit of the powerful computational machinery of INLA. Moreover, it allows to perform multivariate inference on a small set of model parameters.

In the INLA within MCMC context, the {\tt inla()} function can be seen as a device to reduce the dimensionality of the model so that one has to focus only on a smaller subset of parameters $\bm{z}_c$. Implementation is also simpler compared to the one of a full MCMC algorithm.
INLA within MCMC is a computationally intensive and sequential algorithm and might take time to converge. Moreover, the present implementation of the {\tt inla()} program is not optimal in this context as it creates a large amount of temporary files every time the model is run.
\citet{Rubio2019} explore simpler alternatives to the full INLA within MCMC approach useful when fitting complex spatial models. These simple alternatives are based on exploring the posterior of the conditioning parameters $\bm{z}_c$ by using a central composite design or simply fixing their value at posterior mode.

%
%

\section{Discussion}

Since its introduction, INLA has established itself as a powerful tool to perform Bayesian analysis on LGMs.
The associated {\tt R-INLA} package has made INLA a practical and
relatively straight forward tool that has reached practitioners in a
wide range of applied field, see  \citet{inla_review1} for a review of
applications. The {\tt R-INLA} package aims at being as general as
possible within the class of LGM. It provides a large selection of
likelihoods, latent models and priors to choose from, and the
possibility to add some user defined latent models and priors. It is
possible to simultaneously model data from different likelihoods,
replicate and copy parts of the latent fields and several other
features, see \citet{inla_review1, newfeatures} and \url{www.r-inla.org}. Recent developments aim to extend
the scope of INLA by combining it with MCMC techniques.

For some of the end-users, interested only in a sub-class of the possible LGMs, the generality of  {\tt R-INLA} comes at the cost of increased complexity and lack of more specific tools. In the years, a series of add-on packages have been created to improve accessibility for a specific target audience and to provide specialized tools that are mainly relevant for the specific class of models under considerations. Table~\ref{packages_table} collects a list of such add-on packages the authors are aware of. For each package a short description of its purpose is reported together with a reference and a url address for download.

    \begin{tabularx}{\linewidth}{l|X|X|X}
    
     {\bf Package name} & {\bf Purpose} & {\bf Reference}  &{\bf Download} \\\hline\hline
      
      {\tt AnimalINLA}   &  Analysis of  ``animal models''/additive
      genetic models/pedigree based models
      & \citet{Holand2013}& \url{https://folk.ntnu.no/annamaho/AnimalINLA/AnimalINLA_1.4}\\\hline
      
      {\tt ShrinkBayes} &  Shrinkage priors with applications to RNA
      sequencing
     &\citet{VanDeViel2012}&\url{https://github.com/markvdwiel/ShrinkBayes}\\\hline
     
      {\tt meta4diag}  & Bayesian inference for bivariate meta-analysis of diagnostic test studies & \citet{JSSv083i01}&\url{https://cran.r-project.org/web/packages/meta4diag}\\\hline
     
      {\tt BAPC} & Bayesian age-period-cohort models with  focus on projections&
      \citet{riebler2017projecting}& \url{https://rdrr.io/rforge/BAPC/man/BAPC.html}\\\hline
     
      {\tt diseasemapping }   & Formatting of population and case data, calculation of Standardized Incidence Ratios, and fitting the BYM model using INLA. &  \citet{JSSv063i12} & \url{https://rdrr.io/rforge/diseasemapping/}\\\hline
      
      {\tt geostatp }    & Geostatistical modelling facilities using Raster and SpatialPoints objects. Non-Gaussian models are fit using INLA.& \citet{JSSv063i12} & \url{https://cran.r-project.org/web/packages/geostatsp}\\\hline
      
      {\tt excursions }    & Excursion sets, contour credible regions, and
      simultaneous confidence bands & \citet{bolin2015} & \url{https://cran.r-project.org/web/packages/excursions}\\\hline
      
      {\tt INLABRU }    & Model spatial distribution and change from
      ecological survey data & \url{www.inlabru.org} & \url{https://cran.r-project.org/web/packages/inlabru} \\\hline
      
      {\tt INLABMA}    & Spatial Econometrics models using Bayesian
      model averaging and MCMC inla  & \citet{INLABMA}&\url{https://cran.r-project.org/web/packages/INLABMA}\\\hline
      
       {\tt nmaINLA}    &  Performs network meta-analysis using INLA. Includes methods to assess the heterogeneity and inconsistency in the network.  & \citet{Burak_et_al} &

       \url{https://cran.r-project.org/web/packages/nmaINLA}\\\hline
       
       {\tt INLAutils}    &  Utility Functions for INLA: Additional Plots and Support for ggplot2.&&\url{https://rdrr.io/github/timcdlucas/INLAutils}\\\hline
       
       {\tt abn }    &  Modelling Multivariate Data with Additive Bayesian Networks  & &  \url{https://CRAN.R-project.org/package=abn} \\\hline
       {\tt BCEA }    & Bayesian Cost Effectiveness Analysis   & \citet{baio2011}&  \url{ https://CRAN.R-project.org/package=BCEA } \\\hline
       {\tt DClusterm}    &  Model-based methods for the detection of disease clusters using GLMs, GLMMs and zero-inflated models.

 & \citet{10.1007/978-3-030-01584-8_1} & \url{ https://CRAN.R-project.org/package=DClusterm } \\\hline
 
       {\tt PrevMap}    &  Geostatistical Modelling of Spatially Referenced Prevalence Data & \citet{JSSv078i08} & \url{ https://CRAN.R-project.org/package=PrevMap }\\\hline
       
       {\tt SUMMER}    & Provides methods for estimating, projecting, and plotting spatio-temporal under-five mortality rates.  & \citet{mercer2015, } & \url{ https://CRAN.R-project.org/package=SUMMER } \\\hline
       
   {\tt surveillance}    &  Temporal and Spatio-Temporal Modeling and Monitoring of Epidemic Phenomena & \citet{surv1, surv2} & \url{ https://CRAN.R-project.org/package=surveillance} \\\hline
   
 {\tt survHE}    &  Survival Analysis in Health Economic Evaluation & \citet{} & \url{https://CRAN.R-project.org/package=survHE } \\\hline
      
\caption{List of add-on packages build around  {\tt R-INLA}
       for specialized sub-classes of LGMs.     \label{packages_table}
}\\
    \end{tabularx}

\bibliography{references}



\end{document}